\documentclass[a4paper,11pt]{article}
\usepackage{jheppub} 
\usepackage[colorlinks=true,citecolor=blue,linkcolor=blue]{hyperref}
\usepackage[normalem]{ulem}
\usepackage{amsmath,amssymb,mathtools}
\usepackage{epsfig}
\usepackage{graphicx}               
\usepackage{url}
\usepackage{color,colortbl}
\usepackage{slashed}
\usepackage{bm}
\usepackage{comment}
\usepackage{cancel}
\usepackage{gensymb}
\usepackage{multirow}
\usepackage{placeins}
\usepackage[normalem]{ulem}
\usepackage[dvipsnames]{xcolor}
\usepackage{epstopdf}
\usepackage{soul}
\usepackage[capitalise, english]{cleveref}
\usepackage{siunitx}
\usepackage{xspace}
\usepackage{bbold}
\usepackage{jheppub}

\definecolor{red}{rgb}{0.6,.0706,.1373}
\definecolor{blue}{rgb}{0,0.396,0.741}
\newcommand\myshade{80}
\colorlet{mylinkcolor}{violet}
\colorlet{mycitecolor}{violet}
\colorlet{myurlcolor}{violet}

\hypersetup{
  linkcolor  = mylinkcolor!\myshade!black,
  citecolor  = mycitecolor!\myshade!black,
  urlcolor   = myurlcolor!\myshade!black,
  colorlinks = true
}

\allowdisplaybreaks

\newcommand{\be}{\begin{equation}}
\newcommand{\ee}{\end{equation}}
\newcommand{\bea}{\begin{eqnarray}}
\newcommand{\eea}{\end{eqnarray}}

\newcommand{\sscript}[1]{{\scriptscriptstyle \mathrm{#1}}}

\def\beq#1\eeq{\begin{align}#1\end{align}}

\makeatletter
\providecommand*{\diff}%
  {\@ifnextchar^{\DIfF}{\DIfF^{}}}
\def\DIfF^#1{%
  \mathop{\mathrm{\mathstrut d}}%
    \nolimits^{#1}\gobblespace}
\def\gobblespace{%
  \futurelet\diffarg\opspace}
\def\opspace{%
  \let\DiffSpace\!%
  \ifx\diffarg(%
    \let\DiffSpace\relax
  \else
    \ifx\diffarg[%
      \let\DiffSpace\relax
    \else
        \ifx\diffarg\{%
        \let\DiffSpace\relax
      \fi\fi\fi\DiffSpace}

\title{\boldmath 
Anarchic neutrinos from flavor deconstruction: phenomenology of the lepton
sector}

\author[a]{Gino Isidori,}
\emailAdd{gino.isidori@uzh.ch}

\author[b,c]{Paride Paradisi,}
\emailAdd{paride.paradisi@pd.infn.it}

\author[a,b,c]{Andrea Sainaghi,}
\emailAdd{andrea.sainaghi@phd.unipd.it}

\author[b,c]{and Nud\v{z}eim Selimovi\'c\,}
\emailAdd{nudzeim.selimovic@pd.infn.it}

\affiliation[a]{Department of Physics, University of Zurich}
\affiliation[b]{Istituto Nazionale di Fisica Nucleare (INFN), Sezione di Padova, 
Italy}
\affiliation[c]{Dipartimento di Fisica e Astronomia ”Galileo Galilei”, Universit\`a di Padova, Italy}

\abstract{
We investigate the neutrino sector in the framework of flavor deconstruction  with an inverse-seesaw realization. This setup naturally links the hierarchical charged-fermion masses to the anarchic pattern of light-neutrino mixing. We determine the viable parameter space consistent with oscillation data and study the phenomenology of heavy neutral leptons (HNL) and lepton-flavor-violating (LFV) processes. Current bounds from direct HNL searches and LFV decays constrain the right-handed neutrino scale to a few TeV, while future $\mu \to e$ experiments will probe most of the region with $\Lambda \lesssim 10~\text{TeV}$. Among possible realizations, models deconstructing $\mathrm{SU}(2)_\mathrm{L} \times \mathrm{U}(1)_\mathrm{B-L}$ or $\mathrm{SU}(2)_\mathrm{L} \times \mathrm{U}(1)_\mathrm{R} \times \mathrm{U}(1)_\mathrm{B-L}$ are those allowing the lowest deconstruction scale. }

\begin{document}
\maketitle
\flushbottom

\section{Introduction} 
\label{sec:intro}
The flavor deconstruction (FD) hypothesis offers a systematic framework to explain the origin of fermion mass hierarchies. In this approach, the Standard Model (SM) gauge group is extended in the ultraviolet (UV) to a product of flavor non-universal gauge factors acting separately on each generation, with flavor universality emerging at low energies as an accidental symmetry. The hierarchical structure of the Yukawa couplings arises from the sequential breaking of this extended symmetry, under the assumption that the SM Higgs is charged under the gauge group associated with the third generation~\cite{Bordone:2017bld,Allwicher:2020esa,Davighi:2023iks}.

Models based on the FD hypothesis have seen a growing interest in the last few years~\cite{Greljo:2018tuh,Davighi:2023evx,Barbieri:2023qpf,FernandezNavarro:2023rhv,FernandezNavarro:2023hrf,FernandezNavarro:2024hnv,Fuentes-Martin:2020bnh,Fuentes-Martin:2022xnb,Covone:2024elw,Lizana:2024jby,Fuentes-Martin:2024fpx}
because of two main reasons. First,
this mechanism is effective not only in explaining the observed fermion mass hierarchies, but also in delivering at low energies approximate flavor symmetries that ensure the compatibility  of the new dynamics with flavor-changing neutral-current bounds. Moreover, the first layer of non-standard dynamics—associated with the breaking of universality between  third and light generations—can occur at scales as low as a few TeV, while remaining consistent with collider constraints. This makes FD an appealing starting point for models that address the electroweak hierarchy problem through new TeV-scale physics  coupled mainly to the third generation~\cite{Fuentes-Martin:2020bnh,Fuentes-Martin:2022xnb,Covone:2024elw,Lizana:2024jby}. The phenomenology of such constructions has been extensively explored in~\cite{DiLuzio:2018zxy,Fuentes-Martin:2019ign,Fuentes-Martin:2020luw,Fuentes-Martin:2020hvc,Cornella:2021sby,Allwicher:2021ndi,Haisch:2022afh,Crosas:2022quq,Korajac:2023xtv,Davighi:2023xqn,Allwicher:2023aql,Capdevila:2024gki}, focusing primarily on the quark and charged-lepton sectors.

The neutrino sector poses a specific challenge to this framework~\cite{Fuentes-Martin:2020pww,Greljo:2024ovt,FernandezNavarro:2025zmb}. Unlike the strongly hierarchical pattern of quarks and charged leptons, the neutrino mass matrix exhibits large, order-one mixing angles and only mild mass hierarchies, often referred to as neutrino anarchy. Reconciling this behavior with the hierarchical flavor structure predicted by FD requires additional ingredients. As shown in Ref.~\cite{Greljo:2024ovt}, the problem can be resolved with heavy right-handed neutrinos which are not gauge singlets. In this case, gauge quantum numbers can be adjusted to yield, in a natural way, hierarchical charged-fermion masses and an anarchic light neutrino mass matrix. This mechanism can be implemented both via type-I and inverse-seesaw (ISS) realizations. While Ref.~\cite{Greljo:2024ovt} identified the general conditions under which light-neutrino anarchy arises in FD, the associated phenomenology beyond the light neutrino sector has not yet been explored. 

The ISS realization~\cite{Wyler:1982dd,Mohapatra:1986bd} of this mechanism is particularly interesting since it can be reconciled with the assumption that flavor non-universality between third and light generations occurs around the TeV scale. Proceeding this way, one expects (at least some) heavy neutral leptons (HNLs) with masses below the TeV scale, within the reach of current colliders~\cite{Abdullahi:2022jlv,Abada:2023raf}. Moreover,  sizable mixing among the NHLs and SM leptons is also expected, with potentially observable lepton-flavor-violating (LFV) 
rates of charged leptons. 
The aim of this work is to provide a comprehensive analysis of these phenomenological aspects. We study the structure and interactions of light and heavy leptons in this framework, determine the viable parameter space, and identify the most relevant experimental probes. 

The paper is organized as follows. Section~\ref{sec:setup} introduces the neutrino sector of our framework and the parameterization used throughout the analysis. The goal of this Section is to determine the viable parameter range for masses and mixing of the HNLs implied by the observed neutrino mass matrix, but for their overall scale. 
In Section~\ref{sec:pheno}, we analyse the phenomenology of the exotic leptons, including direct HNL searches and impact on LFV observables. 
The implications for model building of the corresponding findings are also briefly discussed. The results are summarised in the Conclusions.

\section{Mass matrices of the neutral leptons} 
\label{sec:setup}

\subsection{Theory framework}
\label{subsec:th_setup}

We assume the existence of $3 \times 3$ neutrino species: three left-handed SM-like $\nu^i_L$, three right-handed $\nu^i_R$, and three left-handed sterile fermions $s^i_L$ that are singlets under the full UV flavor-deconstructed gauge group. The mass Lagrangian is 
\begin{equation}\label{mass lagrangian neutrinos}
    -\mathcal{L}_{\nu}^{\rm mass}\supset \overline{\ell}_L^iY_\nu^{ij} \tilde{H}\nu_R^j + \overline{s}_L^iM_R^{ij} \nu_R^j+\dfrac{1}{2}\overline{s}_L^i\, \mu^{ij} (s_L^c)^j+\rm{h.c.}\,,
\end{equation}
where the sum over the flavor indices ($i,j=1,2,3$) is understood, $\ell_L=(\nu_L,e_L)^T$ denotes the SM lepton doublets,
and $\tilde{H}\equiv i\sigma_2 H^*$ 
where $H$ is the SM Higgs with 
vacuum expectation value (VEV)
$v\equiv \langle H\rangle \approx 246$~GeV. 
The Majorana mass matrix of the sterile fermions  satisfies $\mu^T = \mu$ and we assume the following general hierarchy 
\be
M_R \gg vY_\nu \gg \mu\,.
\ee
Regardless of the  details of the UV completion, the FD hypothesis implies flavor-hierarchical structures for both $Y_\nu$ 
and $M_R$. More precisely, similarly to quark and charged-lepton Yukawa couplings, 
we expect~\cite{Davighi:2023iks}
\begin{equation}
\label{Y-nu parametric}
{\rm diag} (Y_\nu) \sim {\rm diag}( 
\varepsilon_1\varepsilon_2,  \varepsilon_1 , 1)\,,
\end{equation}
with $\varepsilon_1,\varepsilon_2\ll 1$.\footnote{Here ${\rm diag} (Y_v)$ denotes the eigenvalues of 
 $Y_v$ via its singular-value decomposition. }
In most cases,
the hierarchical structure of the Yukawa couplings also implies that the misalignment between $Y_\nu$ and the charged-lepton Yukawa coupling is controlled by unitary matrices close to the identity. 
Concerning  $M_R$, the structure implied by all models discussed in~\cite{Greljo:2024ovt}, 
where the $\nu^i_R$
are charged under a deconstructed gauge group, whereas the $s^i_L$ are singlets, is
\begin{equation}
\label{M_R parametric}
    M_R \sim \Lambda\begin{pmatrix}
        \eta_1\eta_2\, &  \eta_1  &  1  \\  
        \eta_1\eta_2\, &  \eta_1 &  1  \\ 
        \eta_1\eta_2\, &  \eta_1 &  1
    \end{pmatrix}\,,
\end{equation}
where $\eta_1, \eta_2 \ll 1$, $\Lambda \gg v$, and $\mathcal{O}(1)$ factors are omitted.
The anarchy in the light neutrino masses is ensured by a cancellation between the hierarchies in  $Y_\nu$ and $M_R$, while their overall suppression follows from the smallness of $\mu$.

At low energies, the spectrum of the theory includes three SM-like neutrinos with a Majorana mass matrix
\begin{equation}\label{m_nu as mu}
    m_\nu \approx A\mu A^T\,,
\end{equation} 
where $A\equiv v Y_\nu M_R^{-1}$.
Additionally, the spectrum contains three pseudo-Dirac HNLs, with a mass matrix given by
\begin{equation}\label{M_h def}
    M_{\rm HNL} \approx \begin{pmatrix}
        0 & M_R \\ M_R^T & \mu
    \end{pmatrix}\,.
\end{equation} 
The pseudo-Dirac nature of the HNLs stems from the small Majorana mass term in Eq.~\eqref{M_h def}:
for most practical purposes we can neglect $\mu$ and treat this $6\times 6$ matrix as a Dirac matrix with three non-vanishing eigenvalues. Note that, since $M_R$ is hierarchical, we expect these three eigenvalues (i.e.~the three HNL masses) to be of order
\begin{equation}
    M_3 \sim \Lambda\,, \qquad M_2 \sim \Lambda \eta_1\,, \qquad M_1 \sim \Lambda \eta_1\eta_2\,.
    \label{eq:HNL_masses}
\end{equation}

We postpone the discussion about the UV origin of $\mathcal{L}_{\nu}^{\rm mass}$ in Eq.~\eqref{mass lagrangian neutrinos} to Sect.~\ref{eq:UV}. 
The goal of this section is to determine the range of the free parameters   via a 
bottom-up approach,  by looking at current neutrino data. 
In particular, we aim to determine the relative ratios of HNL masses and their mixing with the SM-like leptons. 
Light neutrino masses are insensitive to the overall scale $\Lambda$. The latter will be constrained in 
Sect.~\ref{subsec: Direct searches} and \ref{subsec: LFV pheno}
by looking at HNL phenomenology. 

\subsection{SM interactions and mass-matrix parameterization}
\label{subsec:SM_int}

The Lagrangian relevant for the neutrino sector, written in the unitary gauge, reads
\begin{equation}\label{lagrangian interactions}
    \mathcal{L}_\nu^{\rm int} = -\left(\dfrac{1}{2}\overline{\nu}_L\hat{m}_\nu \nu_L^c+{\rm h.c.}\right)-\overline{n}\hat{M}n
    +\frac{g}{\sqrt{2}}\left(W_\mu^-J^{\mu}_++{\rm h.c.}\right)
    +\frac{g}{2c_W} Z_\mu\, J_Z^\mu\,,
\end{equation}
with the currents $J^{\mu}_+$ and $J_Z^\mu$ given by
 \begin{subequations}
     \begin{align}        &J^{\mu}_+=\overline{e}_L\gamma^\mu U\nu_L+\overline{e}_L\gamma^\mu Vn_L\,,\label{eq:j+}\\
        &J_Z^\mu = \overline{\nu}_L\gamma^\mu (U^\dag U)\,\nu_L+\overline{n}_L\gamma^\mu (V^\dag V)\,n_L+\left(\overline{\nu}_L\gamma^\mu \,(U^\dag V)\,n_L+\rm{h.c.}\right).\label{eq:jZ}
\end{align}
 \end{subequations}
By definition,  the mass matrices $\hat{m}_\nu$ and $\hat{M}$ are diagonal (see App.~\ref{sec: appendix - mass basis diag} for the explicit rotation to the mass basis), 
\begin{equation}
\label{neutrino mass states diagonal}
    \hat{m}_\nu = {\rm diag}(m_{\nu_1},m_{\nu_2},m_{\nu_3})\,,\quad \hat{M} = {\rm diag}(M_{1},M_{2},M_{3}) = \Lambda\times{\rm diag}(\eta_{1}\eta_2,\eta_1,1) \,,
 \end{equation}
while $U$ is the Pontecorvo–Maki–Nakagawa–Sakata (PMNS) matrix, and $V$ is the $3\times 3$ matrix encoding the mixing between charged leptons with HNLs.  In order to study their effects, we utilize the following parameterization
\begin{equation}
\label{mixing matrices parametrization}
    U=\mathcal{N}\left(1-\dfrac{1}{2}WW^\dag\right)\,,\quad V=\mathcal{N}W\,,
\end{equation}
where $\mathcal{N}$ is the  PMNS matrix in the SM limit (NP decoupling limit), and
\begin{equation}\label{W_parametrization}
    W=\mathcal{U}^\dag \hat{A}U_S^\dag\,,\qquad \hat{A} \approx \dfrac{y_\nu v}{\Lambda}\begin{pmatrix}
        \Delta_1\Delta_2 & & \\ & \Delta_1 & \\ & & 1
    \end{pmatrix}\,,\qquad \Delta_i=\dfrac{\varepsilon_i}{\eta_i}\,,
\end{equation}
with $\mathcal{U}$ and $U_S$ being two $3\times 3$ unitary matrices. In this parameterization, $\hat{A}$ can be safely approximated as diagonal, provided that the unitary matrices that diagonalize $M_R$ and $Y_\nu$ on the right are similar. Otherwise, $\hat{A}$ takes a lower-triangular form. Nevertheless, this does not significantly affect the phenomenology, as discussed in App.~\ref{sec: appendix - corr to A}.

Neutrino oscillation experiments provide precise determinations  of the mixing angles and mass-squared differences among the three active neutrino states. However, the absolute ordering of the neutrino masses is still unknown. In the Normal Ordering (NO) scenario, the mass ordering is $m_{\nu_1}<m_{\nu_2}<m_{\nu_3}$, whereas, in the Inverted Ordering (IO), 
we have $m_{\nu_3}<m_{\nu_1}<m_{\nu_2}$. In the standard parameterization, which is independent of the mass ordering,
the PMNS matrix reads
\begin{equation}
    \mathcal{N}=\begin{pmatrix}
            1 & 0 & 0 \\ 0 & \cos{\theta_{23}} & \sin{\theta_{23}} \\ 0 & -\sin{\theta_{23}} & \cos{\theta_{23}}
        \end{pmatrix}\!\begin{pmatrix}
           \cos{\theta_{13}} & 0 & \sin{\theta_{13}}e^{-i\delta_\sscript{CP}} \\ 0 & 1 & 0 \\ -\sin{\theta_{13}}e^{i\delta_\sscript{CP}} & 0 & \cos{\theta_{13}}
        \end{pmatrix}\!\begin{pmatrix}
            \cos{\theta_{12}} & \sin{\theta_{12}} & 0 \\ -\sin{\theta_{12}} & \cos{\theta_{12}} & 0 \\ 0 & 0 & 1
        \end{pmatrix}.
\end{equation} 
In Tab.~\ref{tab:oscillation neutrino data}, we report the current best fit values for neutrino mixing angles and mass differences, defined as $\Delta m^2_{ij}\equiv m^2_{\nu_i}-m^2_{\nu_j}$, both in NO and IO scenarios. Finally, the upper bound on the absolute neutrino mass scale, as 
obtained by the direct kinematic measurements at the KATRIN experiment, reads~\cite{KATRIN:2024cdt}
\begin{equation}
    m_{\nu_e}^{\mathrm{eff}}\equiv \sum_i m_{\nu_i}|\mathcal{N}_{ei}|<0.45\,\rm{eV}\,, \quad\qquad\quad @\quad 90\%\, \text{CL}\,.
\end{equation}

\begin{table}
\renewcommand{\arraystretch}{1.5}
    \centering
    \begin{tabular}{|c|cccc|cc|}
    \hline
        & $\theta_{12}\,(\degree)$ & $\theta_{23}\,(\degree)$ & $\theta_{13}\,(\degree$) & $\delta_\sscript{CP}\,(\degree$) & $\Delta m^2_{21}\,(10^{-5}\,\rm{eV}^2)$ &  $\Delta m^2_{32}\,(10^{-3}\,\rm{eV}^2)$ \\
        \hline
       NO &  $33.68^{+0.73}_{-0.70}$ & $43.3^{+1.0}_{-0.8}$ & $8.56^{+0.11}_{-0.11}$ &  $212^{+26}_{-41}$ & $7.49^{+0.19}_{-0.19}$  & $+2.513^{+0.021}_{-0.019}$\\
       IO &  $33.68^{+0.73}_{-0.70}$ & $48.6^{+0.7}_{-0.9}$ & $8.58^{+0.11}_{-0.11}$ &  $285^{+25}_{-28}$ & $7.49^{+0.19}_{-0.19}$  & $-2.510^{+0.024}_{-0.025}$\\
        \hline
    \end{tabular}
    \caption{Best fit values of the neutrino parameters from oscillation data~\cite{Esteban:2024eli}, both in Normal Ordering (NO) and Inverted Ordering (IO). The errors denote the $68\%$ CL~intervals.}
    \label{tab:oscillation neutrino data}
\end{table}

\subsection{Range of the parameters}
\label{subsec:range_of_param}

The parameterization introduced in Eqs.~\eqref{mixing matrices parametrization} and~\eqref{W_parametrization} allows the relevant phenomenology to be described in terms of twelve main parameters:  three eigenvalues of the neutrino Yukawa matrix $Y_\nu$;
the three eigenvalues of  $M_R$  (or, equivalently, the scale $\Lambda$ and the ratios $\Delta_{1,2}$); two sets of 
three mixing angles each, $\alpha_i$ and $\beta_j$, through which 
the unitary matrices $\mathcal{U}$ and $U_S$ are parameterized 
as follows
\begin{align}
        \mathcal{U}&=\begin{pmatrix}
            1 & 0 & 0 \\ 0 & \cos{\alpha_1} & \sin{\alpha_1} \\ 0 & -\sin{\alpha_1} & \cos{\alpha_1}
        \end{pmatrix}\begin{pmatrix}
           \cos{\alpha_2} & 0 & \sin{\alpha_2} \\ 0 & 1 & 0 \\ -\sin{\alpha_2} & 0 & \cos{\alpha_2}
        \end{pmatrix}\begin{pmatrix}
            \cos{\alpha_3} & \sin{\alpha_3} & 0 \\ -\sin{\alpha_3} & \cos{\alpha_3} & 0 \\ 0 & 0 & 1
        \end{pmatrix}\,,\\
        U_S&=\begin{pmatrix}
            1 & 0 & 0 \\ 0 & \cos{\beta_1} & \sin{\beta_1} \\ 0 & -\sin{\beta_1} & \cos{\beta_1}\,\,
        \end{pmatrix}\begin{pmatrix}
           \cos{\beta_2} & 0 & \sin{\beta_2} \\ 0 & 1 & 0 \\ -\sin{\beta_2} & 0 & \cos{\beta_2}
        \end{pmatrix}\begin{pmatrix}
            \cos{\beta_3} & \sin{\beta_3} & 0 \\ -\sin{\beta_3} & \cos{\beta_3} & 0 \\ 0 & 0 & 1\,
        \end{pmatrix}\,.
\end{align}
In principle, $\mathcal{U}$ and $U_S$ could be complex; however, barring fine-tuned configurations, the corresponding phases play a negligible role in our discussion and will be ignored. Finally, key ingredients are the entries of the Majorana mass matrix $\mu$ for the sterile fermions. However, as we discuss below, these can be completely determined in terms of the other  parameters and the observed light neutrinos mass matrix, under the assumption that $\mu$ is anarchic.\vspace{0.2cm}

\noindent The reference values, or allowed ranges, for the  parameters listed above are the following: 

\paragraph*{Eigenvalues of $\bm{Y_\nu}$.} The largest eigenvalue, $y_\nu$, is expected to be of order one, while the other two eigenvalues are suppressed by the small parameters $\varepsilon_1, \varepsilon_2 \ll 1$. Motivated by the observed hierarchy in the charged lepton Yukawa matrix $Y_e$, we expect
\begin{equation}
\varepsilon_1 \sim \frac{m_\mu}{m_\tau} \approx 0.06\,, \qquad \varepsilon_2 \sim \frac{m_e}{m_\mu} \approx 0.005\,,
\label{epsilon_values}
\end{equation}
where the above values are representative of the expected order of magnitude.

\paragraph*{Mixing angles in $\bm{\mathcal{U}}$.}
 The three angles $\alpha_i$  can, in principle, vary within the full range $[0, 2\pi]$. However, in flavor-deconstructed models, $\mathcal{U} = \mathcal{N} + \mathcal{O}(\varepsilon_1, \varepsilon_2)$ (see App.~\ref{sec: appendix - choice of param}), implying that the $\alpha_i$ are expected to be close to the PMNS mixing angles, reported in Tab.~\ref{tab:oscillation neutrino data}. Given that $\varepsilon_1, \varepsilon_2 \lesssim 0.1$, a plausible ranges for the $\alpha_i$ is
\begin{equation}\label{alpha i region}
\alpha_1 \in [\theta_{23} - 0.1, \theta_{23} + 0.1]\,,\quad
\alpha_2 \in [\theta_{13} - 0.1, \theta_{13} + 0.1]\,,\quad
\alpha_3 \in [\theta_{12} - 0.1, \theta_{12} + 0.1]\,.
\end{equation}
In the limit where these angles approach the PMNS values, their impact on the phenomenology becomes negligible. As a result, the effective number of parameters controlling the relevant observables is reduced to nine.

\paragraph*{Mixing angles in $\bm{U_S}$.} The three angles $\beta_i$  can also vary in the full range $[0, 2\pi]$. As shown in App.~\ref{sec: appendix - choice of param}, this matrix originates purely from UV dynamics involving only the singlet fermions $s_L$ of the ISS, and is thus unconstrained by low-energy observables. The only constraint is that all three angles should not simultaneously be closer to $0$ or $\pi$ than a quantity $\mathcal{O}(\eta_i)$, as this would inhibit the generation of the anarchic structure in $M_R$, see Eq.~\eqref{M_R parametric}. Since such a fine-tuned scenario departures from our starting assumptions, we neglect it in the following.
 
\paragraph*{Range of $\bm{\Delta_{1,2}}$.}
 Having chosen reference values for the $Y_\nu$ eigenvalues, and determined the range for $\alpha_i$ and $\beta_i$, we proceed determining the allowed range for the parameters controlling the eigenvalues of 
 $M_R$, namely $\Lambda$ and ratios $\Delta_i = \varepsilon_i / \eta_i$ $(i = 1, 2)$. 
 These parameters play a central role in our analysis since they control the spectrum of the HNLs.
 Looking only at light neutrino data, in the absence of independent information about the overall normalization of $\mu$, it is not possible to determine the value of $\Lambda$. On the other hand, we can constrain the range of $\Delta_{1,2}$.

The strategy we follow to determine this range is to reproduce the observed light neutrino spectrum under the assumption that the sterile Majorana mass matrix $\mu$ is anarchic. The latter can be expressed as (see App.~\ref{sec: appendix - choice of param} for a derivation)
\begin{equation}\label{mu from mnu}
    \mu =U_S\left[\hat{A}^{-1}\mathcal{U}\,\hat{m}_\nu\,\mathcal{U}^T(\hat{A}^{-1})^T\right]U_S^T\,,
\end{equation}
where $\hat{m}_\nu$ is the active neutrino mass matrix in Eq.~\eqref{neutrino mass states diagonal} and, as already stated, $\mathcal{U}$ is expected to be close to the PMNS matrix.

On general grounds, we define an \textit{anarchic matrix} to have real entries (for simplicity) that are randomly distributed in the interval $[1,10]$, with randomly assigned signs ($\pm$), and multiplied by an overall scale. 
In principle, given an anarchic structure for $\mu$, one can constrain the parameter space of the $\Delta_i$ to reproduce the observed neutrino data. However, this approach is computationally inefficient, and therefore, we prefer to take the neutrino oscillation data as an input, rather than trying to reproduce it as output. By using the known structure of the light neutrino mass matrix, we thus constrain the parameter space of the $\Delta_i$, such that the resulting matrix $\mu$ approximately satisfies the anarchic condition. 

\begin{figure}[t]
    \centering
    \includegraphics[width=0.6\linewidth]{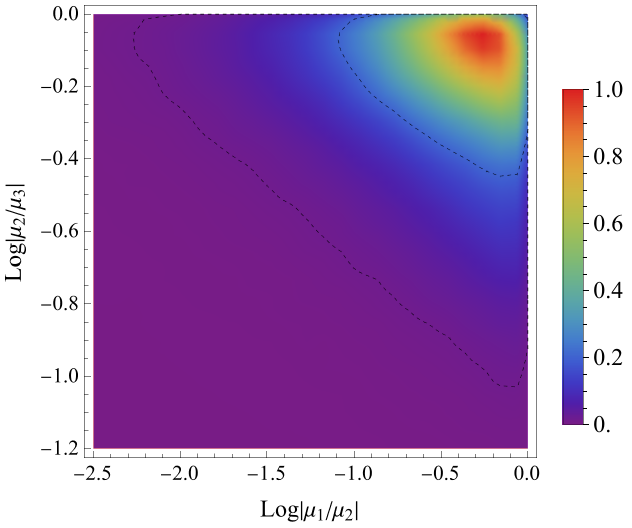}
    \caption{Probability distributions of the logarithmic ratios of the eigenvalues $|\mu_1|<|\mu_2|<|\mu_3|$ of an anarchic random matrix $\mu$. The dashed-black lines contain $68\%$ and $95\%$ of the distribution.}
    \label{fig:mu anarchic stat distr}
\end{figure}
 
This approach is limited by the randomness introduced by the mixing angles $\beta_i$ in $U_S$, which can partially distort the desired anarchic structure of $\mu$. However, the eigenvalues of $\mu$ are invariant under unitary rotations and thus independent of $U_S$. Denoting the ordered eigenvalues by $|\mu_1|<|\mu_2|<|\mu_3|$, we analyze their statistical distribution for randomly generated anarchic matrices. Since we are not concerned with the overall normalization of $\mu$, we focus on the two independent ratios $|\mu_1|/|\mu_2|$ and $|\mu_2|/|\mu_3|$. The resulting two-dimensional statistical distribution is shown in Fig.~\ref{fig:mu anarchic stat distr}. To better visualize the regions where the ratios approach zero, we present the distribution on a logarithmic scale. 

Moreover, to quantitatively assess the preferred values of the parameters $\Delta_i$, we adopt the following procedure:
\begin{itemize}
    \item Generate an ensemble of $N=100$ neutrino mass spectra consistent with current experimental constraints, varying the absolute neutrino mass scale and considering both normal and inverted mass orderings.
    \item For fixed values of $\Delta_{1,2}$, compute the matrix $\mu$ using Eq.~\eqref{mu from mnu}, drawing the parameters $\alpha_i$ randomly distributed over the domain specified in Eq.~\eqref{alpha i region}. We then impose the condition that the eigenvalues of $\mu$ lie within the region encompassing the $95\%$ most probable configurations of the statistical distribution shown in Fig.~\ref{fig:mu anarchic stat distr}.
    \item Repeat the procedure scanning over the values of $\Delta_1$ and $\Delta_2$, thereby identifying the parameter regions that yield a statistically consistent anarchic $\mu$.
\end{itemize} 

\begin{figure}[t]
    \centering
        \includegraphics[width=0.3 \linewidth]{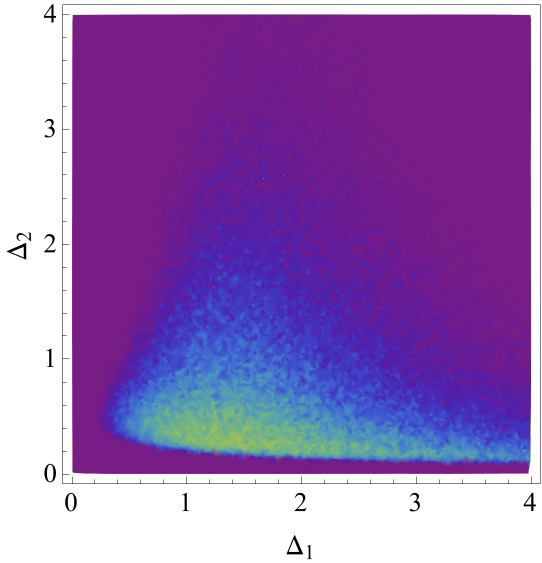} \includegraphics[width=0.3  \linewidth]{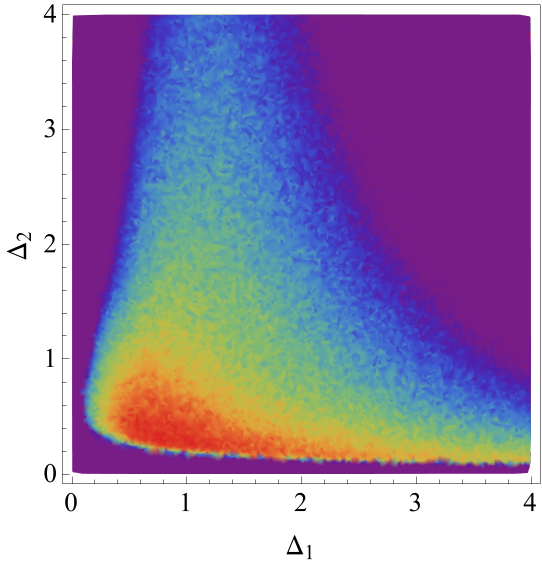} 
        \centering{\includegraphics[width=0.3 \linewidth]{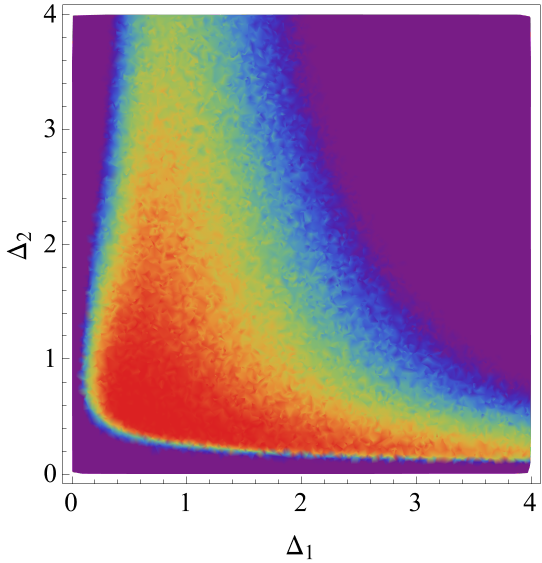}}
        \includegraphics[width=1cm]{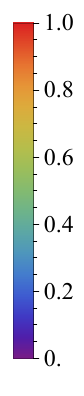}
    \caption{Probability for the parameter $\Delta_i$ to satisfy our anarchic criterion on $\mu$ at $95\%$ CL,
    assuming Normal Ordering. The three plots refer to
    different values of the absolute neutrino mass scale:  $m_{\nu_1}\in [0.1,1]$~meV (left panel),
     $m_{\nu_1}\in [1,10]$~meV (central panel), 
     and $m_{\nu_1}\in [10,50]$~meV (right panel).}
    \label{fig:NO delta range}
\end{figure}

\begin{figure}[t]
    \begin{minipage}[r]{0.85\textwidth}
        \hspace{0.2\textwidth}
        \includegraphics[width=0.35\linewidth]{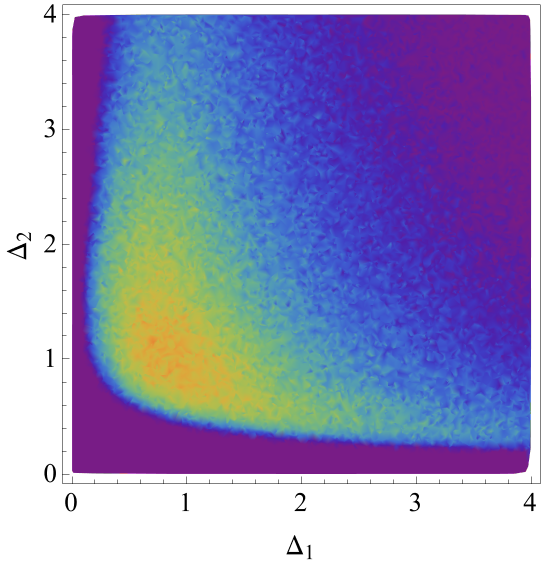}\quad \includegraphics[width=0.35\linewidth]{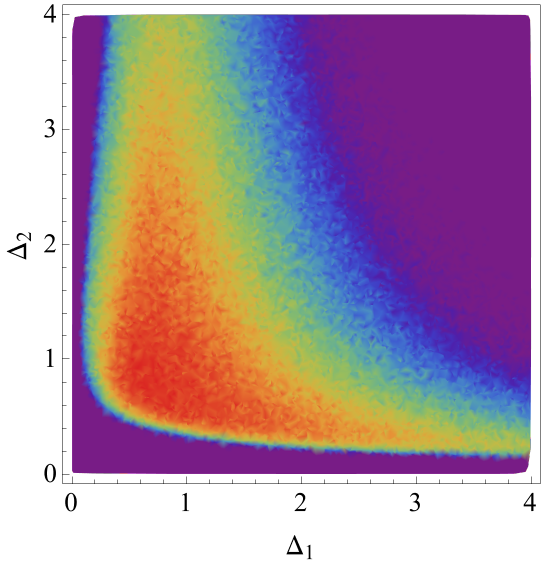}
    \end{minipage}
    \begin{minipage}[l]{0.1\textwidth}
        \includegraphics[width=1cm]{Plots/legend_anarchic.pdf}
    \end{minipage}
    \caption{Probability for the parameter $\Delta_i$ to satisfy our anarchic criterion on $\mu$ at $95\%$ CL,
     assuming Inverted Ordering. The two plots refer to
    different values of the absolute neutrino mass scale:  $m_{\nu_3} \in [1,10]$~meV (left panel), and
     $m_{\nu_1}\in [10,50]$~meV (right panel).}
    \label{fig:IO delta range}
\end{figure}

This procedure enables us to identify the ranges of the    $\Delta_i$ that are consistent with different configurations of the neutrino mass spectrum. The results are shown in Fig.~\ref{fig:NO delta range}~(NO case) 
and Fig.~\ref{fig:IO delta range}~(IO case), for different values of the lightest neutrino mass. 
As expected, the $\Delta_i$ are of $\mathcal{O}(1)$; however, 
their allowed range has a non-trivial dependence on the absolute neutrino mass scale. In particular, 
 we find that no viable region exists for $m_{\nu_1}<0.1$\,meV
 and   $m_{\nu_3}<1$\,meV for NO and IO, respectively.
 More generally, neutrino masses $m_{\nu}\gtrsim 10$ meV tend to admit broader regions of viable $\Delta_i$. Moreover,   the normal ordering scenario typically allows for slightly smaller values of $\Delta_i$ compared to the inverted ordering case.
 
\section{Phenomenology of standard and exotic leptons } 
\label{sec:pheno}

Having determined the allowed range for the $\Delta_i$ parameters,
in the  following, we examine the most relevant constraints on HNL phenomenology, spanning from direct searches to their indirect effects on LFV processes, in order to determine constraints on $\Lambda$.

\subsection{Direct searches of HNL}
\label{subsec: Direct searches}

As discussed in Sect.~\ref{subsec:th_setup}, the FD of the ISS mechanism comes with the prediction of three HNLs with hierarchical masses. The lightest one, $n_1$, could have a mass $M_1$ low enough to be detected in direct searches at colliders. For instance, if $M_1< M_{W,Z}$, a lower bound on   $\Lambda$ can be inferred through the tree-level processes $Z\rightarrow \nu n_1$ and $W\rightarrow \ell n_1$. 
These bounds depend on the size of  $|V_{\ell1}|$, which controls the mixing between $n_1$ and the charged lepton   ($\ell=e,\mu,\tau$), see Eqs.~\eqref{lagrangian interactions} and~\eqref{mixing matrices parametrization}.
The current limits on this mixing  as a function of the HNL mass can be found in Refs.~\cite{Fernandez-Martinez:2023phj,Abdullahi:2022jlv}. 

\begin{figure}[t]
    \centering
        \includegraphics[width=0.3 \linewidth]{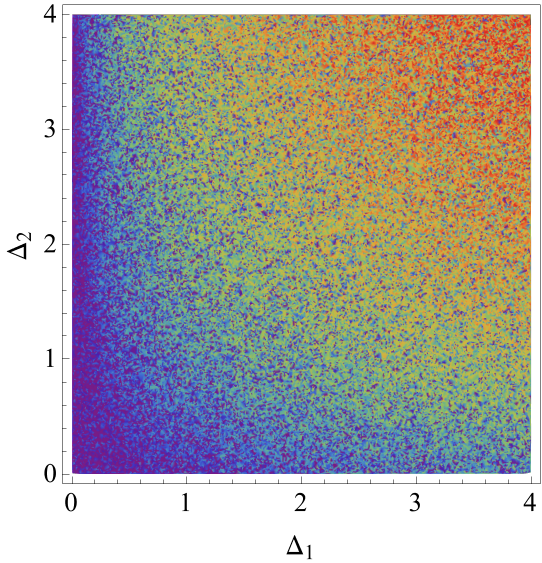} \includegraphics[width=0.3  \linewidth]{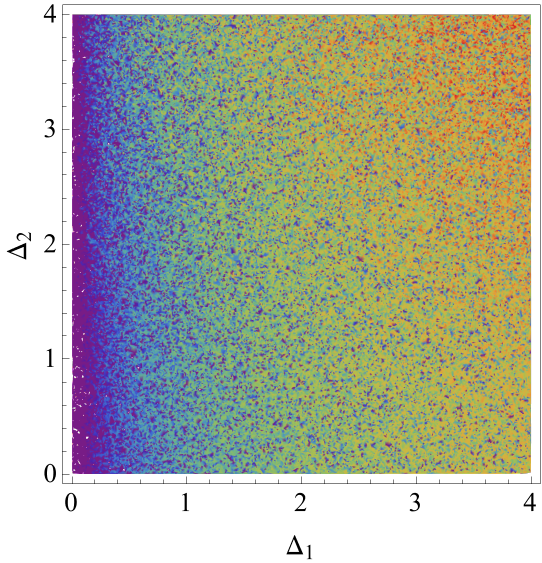} 
        \centering{\includegraphics[width=0.3 \linewidth]{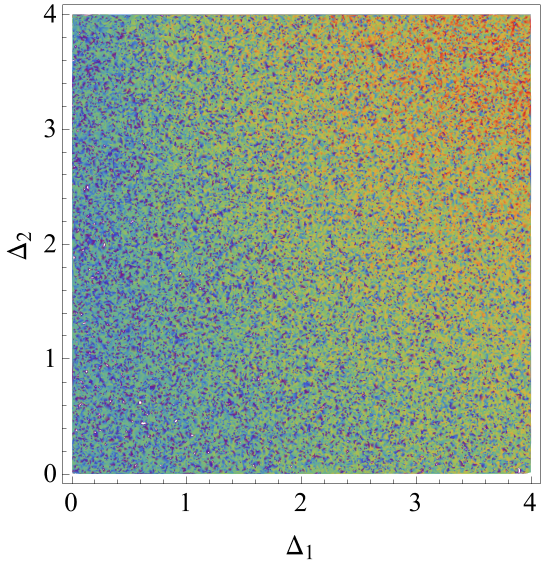}}
        \includegraphics[width=0.9cm]{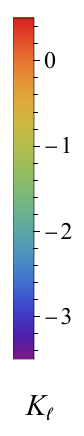}
    \caption{Plots showing the value of the parameter $K_\ell$ controlling the mixing between the lightest HNL and the SM charged lepton $\ell=e$ (left panel), $\mu$ (central panel) and $\tau$ (right panel), as a function of the $\Delta_i$.}
    \label{fig:K dependence - mixing}
\end{figure}

To translate these bounds into constraints on $\Lambda$, 
it is useful to define the following dimensionless quantity
\begin{equation}\label{Kell def}
    K_\ell\equiv \mathrm{Log}\left[|V_{\ell1}|^2|y_\nu|^{-2}\left(\dfrac{\Lambda}{\mathrm{TeV}}\right)^2\right]\,,
\end{equation}
which depends only on $\Delta_i$ and the $\beta_i$ angles through $V_{\ell 1}\sim y_\nu\, v/\Lambda$. Though not flavor universal, the coefficients $K_\ell$ are of the same order for each flavor, see Fig.~\ref{fig:K dependence - mixing}. By inverting Eq.~\eqref{Kell def}, we find the following bound on the NP scale
\begin{equation}
\label{eq:ZWbound}
    \Lambda > y_\nu \,\sqrt{\dfrac{10^{-5}}{|V^{\mathrm{mix}}_{\ell n_1}|^2_{\mathrm{exp}}}}~
    10^{\frac{5+K_\ell}{2}}\,\,\,\mathrm{TeV}\,,
\end{equation}
where $|V^{\mathrm{mix}}_{\ell n_1}|^2_{\mathrm{exp}}$ refers to the current experimental bound on the mixing angles. The strongest constraint on $\Lambda$ from Eq.~\eqref{eq:ZWbound}, for masses in the range $15$~GeV$\,\lesssim M_1\lesssim 90$~GeV, is currently inferred from DELPHI and LHC limits on exotic $Z$ and $W$ decays, respectively~\cite{Fernandez-Martinez:2023phj,Abdullahi:2022jlv}. The bound reads
\begin{equation}
    \Lambda\gtrsim 20\,y_\nu\,\,\mathrm{TeV}\qquad \text{for}\qquad \Delta_i<1\,.
    \label{eq:Lambda_bound}
\end{equation}
This bound can be satisfied for $\Lambda\sim$~few TeV only if $y_\nu\sim 
\mathcal{O}(0.1)$. This condition implies, in turn, a strong suppression for LFV processes (see Sect.~\ref{subsec: LFV pheno}).
For masses $M_1>M_Z$, the current experimental contraints on the mixing angles $|V^{\mathrm{mix}}_{\ell n_1}|^2_{\mathrm{exp}}$ are much weaker~\cite{Abdullahi:2022jlv}, and we find the following bound
\begin{equation}
    \Lambda\gtrsim y_\nu\,\,\mathrm{TeV}\qquad \text{for}\qquad \Delta_i=\mathcal{O}(1)\,.
    \label{eq:Lambda_bound_Mn big}
\end{equation}

In addition, the relation between $M_1$ and $\Lambda$ imposes itself a further bound on the NP scale. Recalling that $M_1= \eta_1\eta_2 \Lambda$ with $\eta_i= \varepsilon_i/\Delta_i$, see Eqs.~\eqref{neutrino mass states diagonal} and \eqref{W_parametrization}, we find  
\begin{equation}\label{lambda bounds}
    \Lambda \approx10\left(\dfrac{M_1}{M_Z}\right)\left(\dfrac{\Delta_1\Delta_2}{0.4}\right)\left(\dfrac{m_\mu^2/m_\tau^2}{\varepsilon_1\varepsilon_2}\right)\,{\rm TeV}\,.
\end{equation}
Remarkably, at present Eq.~\eqref{lambda bounds} provides a better bound than Eq.~\eqref{eq:Lambda_bound_Mn big} on the NP scale $\Lambda$ in the case $M_1>M_Z$, allowing $\Lambda\sim$~few TeV for reasonable choices of the model parameters.
The situation will change in the future with dedicated searches of HNLs, with mass around or above $M_Z$, at future  $e^+e^-$ colliders, which are expected to improve current bounds by more than one order of magnitude.

\subsection{Low-energy constraints}
\label{subsec: LFV pheno}

\begin{table}
    \centering
    \scalebox{0.94}{
    \begin{tabular}{|c|c|c|}
    \hline
        Observable & Current bound & Future sensitivity  \\
        \hline
        \hline
         $\mathcal{B}(\mu\rightarrow e\gamma)$ & $<1.5\times 10^{-13}$ (MEG II \cite{MEGII:2025gzr}) & $6\times 10^{-14}$ (MEG II \cite{MEGII:2025gzr}) \\
         $\mathcal{B}(\mu\rightarrow eee)$ & $<1.0\times 10^{-12}$ (SINDRUM \cite{SINDRUM:1987nra}) & $10^{-15}\div 10^{-16}$ (Mu3e \cite{Blondel:2013ia}) \\
         $\mathrm{Cr}(\mu-e,\rm{N})$ & $<7\times 10^{-13}$ (Au, SINDRUM II \cite{SINDRUMII:2006dvw}) & $2.6\times  10^{-17}$ (Al, COMET \cite{Krikler:2015msn}) \\
         \hline
         $\mathcal{B}(\tau\rightarrow e\gamma)$ & $<3.3\times 10^{-8}$ (BaBar \cite{BaBar:2009hkt}) & $3\times 10^{-9}$ (Belle II \cite{Belle-II:2018jsg}) \\
         $\mathcal{B}(\tau\rightarrow \mu\gamma)$ & $<4.4\times 10^{-8}$ (BaBar \cite{BaBar:2009hkt}) & $10^{-9}$ (Belle II \cite{Belle-II:2018jsg}) \\
         $\mathcal{B}(\tau\rightarrow eee)$ & $<2.7\times 10^{-8}$ (Belle \cite{Hayasaka:2010np}) & $5\times10^{-10}$ (Belle II \cite{Belle-II:2018jsg}) \\
         $\mathcal{B}(\tau\rightarrow \mu\mu\mu)$ & $<3.3\times 10^{-8}$ (Belle \cite{Hayasaka:2010np}) & $5\times10^{-11}$ (FCC-ee \cite{FCC:2018byv}) \\
         $\mathcal{B}(\tau^-\rightarrow e^-\mu^-\mu^+)$ & $<2.7\times 10^{-8}$ (Belle \cite{Hayasaka:2010np}) & $5\times10^{-10}$ (Belle II \cite{Belle-II:2018jsg}) \\
         $\mathcal{B}(\tau^-\rightarrow e^-\mu^-e^+)$ & $<1.8\times 10^{-8}$ (Belle \cite{Hayasaka:2010np}) & $5\times10^{-10}$ (Belle II \cite{Belle-II:2018jsg}) \\
         $\mathcal{B}(\tau^-\rightarrow e^-e^-\mu^+)$ & $<1.5\times 10^{-8}$ (Belle \cite{Hayasaka:2010np}) & $5\times10^{-10}$ (Belle II \cite{Belle-II:2018jsg}) \\
         $\mathcal{B}(\tau^-\rightarrow \mu^-\mu^-e^+)$ & $<1.7\times 10^{-8}$ (Belle \cite{Hayasaka:2010np}) & $5\times10^{-10}$ (Belle II \cite{Belle-II:2018jsg}) \\
         \hline
         $\mathcal{B}(Z\rightarrow e\mu)$ & $<4.2\times 10^{-7}$ (ATLAS \cite{ATLAS:2021bdj}) & $\mathcal{O}(10^{-10})$ (FCC-ee \cite{FCC:2018byv}) \\
         $\mathcal{B}(Z\rightarrow e\tau)$ & $<4.1\times 10^{-6}$ (ATLAS \cite{ATLAS:2021bdj}) & $\mathcal{O}(10^{-10})$ (FCC-ee \cite{FCC:2018byv}) \\
         $\mathcal{B}(Z\rightarrow \mu\tau)$ & $<5.3\times 10^{-6}$ (ATLAS \cite{ATLAS:2021bdj}) & $\mathcal{O}(10^{-10})$ (FCC-ee \cite{FCC:2018byv}) \\
         \hline
    \end{tabular}
    }
    \caption{Current and future experimental bounds on the LFV observables considered in this work.}
    \label{tab:main LFV bounds}
\end{table}

The mixing of the HNLs with the SM-like leptons is responsible for a series of non-standard effects that we discuss in detail below. These include LFV decays of $\mu$ and $\tau$ leptons, LFV $Z$ decays, and deviations from unitarity in the PMNS matrix. Among them, $\mu\to e$ transitions are those posing 
the most stringent constraints on the model parameter space. 
Current experimental bounds and future sensitivities on the various LFV rates are  summarized in Tab.~\ref{tab:main LFV bounds}. 

\paragraph*{LFV $\bm{\mu}$ decays.}
The  $\mu\rightarrow e\gamma$ branching fraction can be written as
\begin{equation}
    \mathcal{B}(\mu\rightarrow e\gamma)=\dfrac{3\alpha}{2\pi}|\delta_\nu|^2\,,\qquad\qquad \delta_\nu=\sum_k V_{ek}V_{\mu k}^*\, G_\gamma\left(\dfrac{M_k^2}{M_W^2}\right)\,,
    \label{eq:Br_mutoegamma}
\end{equation}
where $G_\gamma(x)$ is a known loop function~\cite{Ilakovac:1994kj}. It turns out that $\mathcal{B}(\mu\rightarrow e\gamma)$ depends  smoothly on $\varepsilon_i$, through the dependence on $M_k$. On the other hand, given $\mathcal{U}\approx \mathcal{N}$, which implies $V\approx \hat{A}U_S$ (see Sect.~\ref{subsec:range_of_param}), we can rewrite $\delta_\nu$ as
\begin{equation}\label{loop form factor}
    \delta_\nu\approx \dfrac{y_\nu^2v^2}{\Lambda^2}\,
    \bar \Delta^3\, \sum_k(U_S)_{ek}(U_S^*)_{\mu k}\, G_\gamma\left(\dfrac{M_k^2}{M_W^2}\right)\,,
    \qquad \bar \Delta = (\Delta_1^2\Delta_2)^{1/3}\,,
\end{equation}
factoring out the dependence on the $\Delta_i$. As we have seen, the latter are expected to be $\mathcal{O}(1)$; however, minor deviations from unity can lead to large deviations in the rates. Indeed, since $\mathcal{B}(\mu\rightarrow e\gamma) \sim |\bar \Delta|^6$, moving from $\bar \Delta=1.0$ to $\bar \Delta=0.5$ leads to a  $10^{-2}$ suppression in the rate. The expression (\ref{loop form factor}) is insensitive to the precise values of the $\alpha_i$, within the range identified in Sect.~\ref{subsec:range_of_param} (it remains a good approximation for deviations from the PMNS values up to 0.2~rad). On the other hand, varying the $\beta_i$ within their allowed range  produces a further variation on the LFV rate (at fixed $\Delta_i$ and $\Lambda$) which can easily exceed two orders of magnitude.

\begin{figure}
    \centering
    \includegraphics[width=0.8\linewidth]{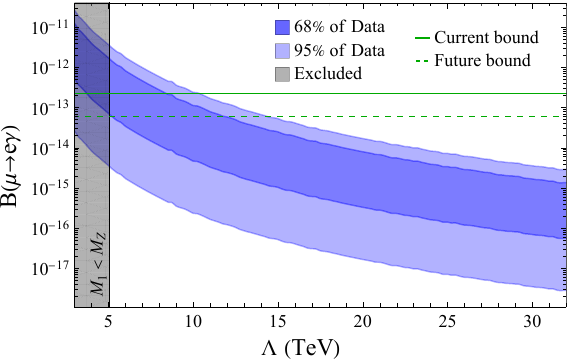}
    \caption{Prediction of $\mathcal{B}(\mu\rightarrow e\gamma)$ as a function of the NP scale $\Lambda$. The prediction is obtained by varying 
    $\alpha_i$ and $\beta_i$   in the range discussed in Sect.~\ref{subsec:range_of_param} and setting $\Delta_1=0.45$, $\Delta_2=0.3$, $y_\nu=1$, $\varepsilon_1=0.06$ and $\varepsilon_2=0.04$. The dark-blue (light-blue) band corresponds to the $68\%$ ($95\%$) of the predicted values  for a given $\Lambda$. Green lines refer to the present and future experimental limits. The vertical gray band denotes the limit from direct HNL searches at colliders.}
    \label{fig:Br-mutoegamma}
\end{figure}

Fig. \ref{fig:Br-mutoegamma} shows $\mathcal{B}(\mu\rightarrow e\gamma)$ as a function of $\Lambda$, for fixed values of $\Delta_1=0.45$, $\Delta_2=0.3$, $\varepsilon_1=0.06$, $\varepsilon_2=0.04$, $y_\nu=1$, with $\alpha_i$ and $\beta_i$  varied in the range discussed in Sect.~\ref{subsec:range_of_param}. As can be seen, $\mathcal{B}(\mu\rightarrow e\gamma)$ satisfies the current experimental bound even for $\Lambda$ around a few TeV. The large span of the rate at fixed $\Lambda$ is driven by the unknown values of the mixing angles $\beta_i$. Assuming a flat distribution for the $\beta_i$, the corresponding range for  $\mathcal{B}(\mu\rightarrow e\gamma)$ covers three orders of magnitude at $95\%$~CL. 

Putting all the ingredients together and isolating the leading parametric dependence, we can write 
\begin{equation}
\label{mueGamma parametric numeric}
    \mathcal{B}(\mu\rightarrow e\gamma) = 
    \kappa_{\mu e \gamma} 
 \times 10^{-13} \times y^4_\nu \left(\dfrac{\bar \Delta}{0.5}\right)^6 \left(\dfrac{10 \,\mathrm{TeV}}{\Lambda}\right)^4\,,
\end{equation}
where $\kappa_{\mu e \gamma}$ is a numerical factor that encodes the variation of the other parameters and, at $95\%$~CL, lies in the range $[10^{-2}, 10]$.

\begin{figure}
    \centering
    \includegraphics[width=0.48\linewidth]{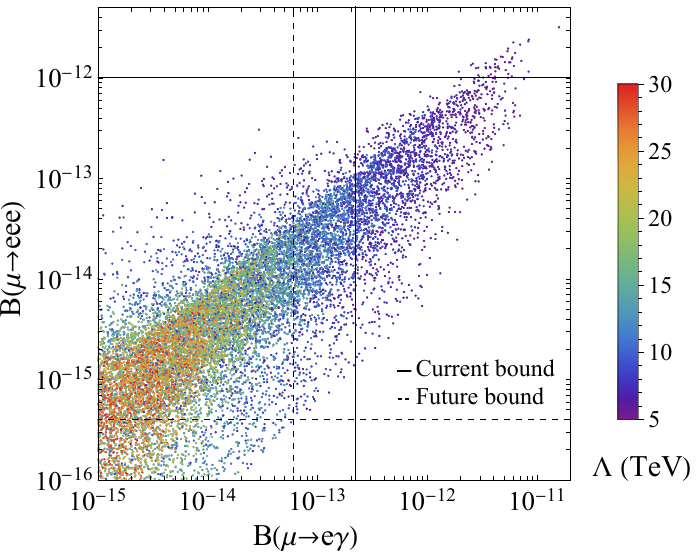}\includegraphics[width=0.48\linewidth]{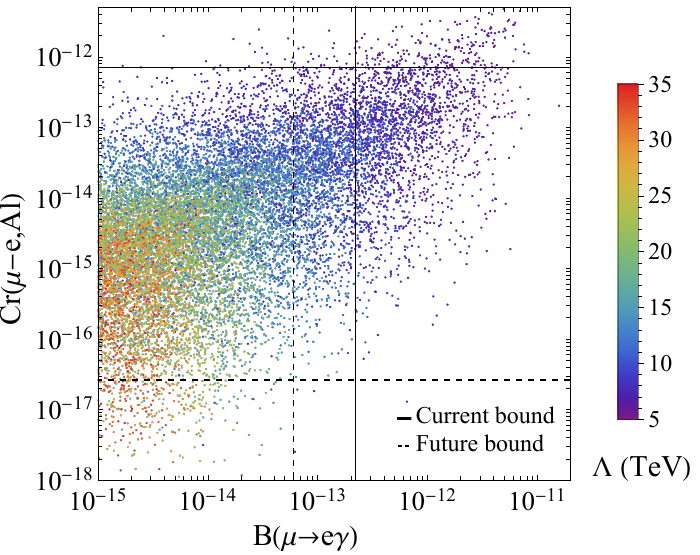}\\ \includegraphics[width=0.48\linewidth]{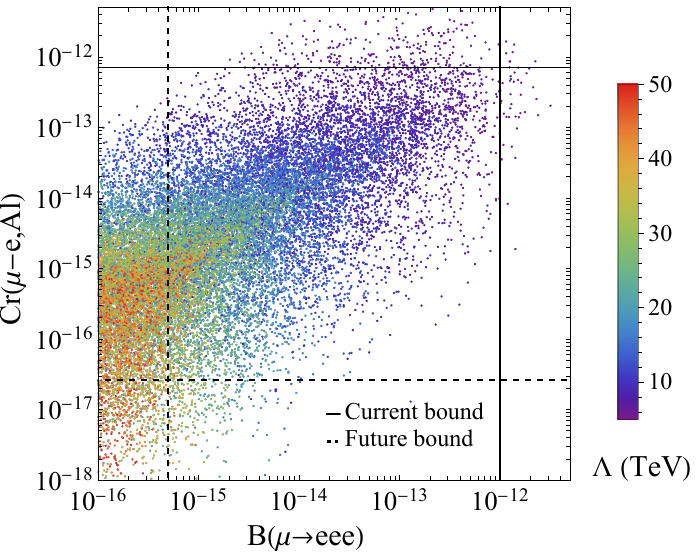}
    \caption{Correlations among the three dominant LFV processes. Continuous (dashed) lines denote the current (future) experimental bounds. The predictions are obtained varying $\alpha_i$ and $\beta_i$   in the range discussed in Sect.~\ref{subsec:range_of_param}, and setting  $\Delta_1=0.5$, $\Delta_2=0.5$, $y_\nu=1$, $\varepsilon_1=0.06$ and $\varepsilon_2=0.02$. Here $\Lambda$ varies between $5$ TeV and $50$ TeV (as indicated by the color scale).}
    \label{fig:correlations_1}
\end{figure}

Proceeding in a similar manner, we analyse the expectations for $\mu\rightarrow eee$ and $\mu-e$ conversion in nuclei. The theoretical predictions of these processes can be found, for instance, in \cite{Ilakovac:1994kj,Kitano:2002mt}. Not surprisingly, the leading parametric dependence from the model parameters is the same one found in the $\mu\rightarrow e\gamma$ case. In Fig.~\ref{fig:correlations_1}, we show the correlations among the three $\mu\to e$ processes  for fixed values of $\Delta_1=0.5$, $\Delta_2=0.5$, $\varepsilon_1=0.06$, $\varepsilon_2=0.02$, $y_\nu=1$, varying $\Lambda$ in the range $[5,50]$ TeV and the mixing angles  in the intervals discussed in Sect.~\ref{subsec:range_of_param}. Currently, $\mu\rightarrow e\gamma$ is the process that constrains this scenario the most;  however, this will change with future projections. 

The analogous expressions of Eq.~\eqref{mueGamma parametric numeric} for $\mathcal{B}(\mu\rightarrow eee)$ and $\mathrm{Cr}(\mu-e,\rm{Al})$ read
\begin{gather}
     \mathcal{B}(\mu\rightarrow eee) = \kappa_{\mu3e} \times 10^{-14}
      \times y^4_\nu \left(\dfrac{\bar \Delta}{0.5}\right)^6 
      \left(\dfrac{10 \,\mathrm{TeV}}{\Lambda}\right)^4\,,\\
    \mathrm{Cr}(\mu-e,\mathrm{Al}) = \kappa_{\mu2e} \times 10^{-13}
    \times y^4_\nu \left(\dfrac{\bar \Delta}{0.5}\right)^6 
    \left(\dfrac{10 \,\mathrm{TeV}}{\Lambda}\right)^4\,,
\end{gather}
where $\kappa_{\mu 3e} \in [10^{-2},10]$ and $\kappa_{\mu2e } \in [10^{-1},10]$.
These have to be compared with the current and future sensitivities summarized in Tab.~\ref{tab:main LFV bounds}. 

\medskip

In general, within FD models, LFV transitions receive additional contributions from the exchange of the massive vector bosons resulting from the 
deconstruction of the gauge group.  In particular, all models feature a 
flavor non-universal $Z^\prime$ which contributes, at the tree-level, to four-fermion processes such as $\mu\rightarrow eee$ and $\mu-e$ conversion
(see App.~\ref{sec:Gauge and yukawa}). In the case of $\mu\rightarrow eee$, we expect the following parametric ratio between tree-level
$Z^\prime$ contribution and the loop-induced rate related to the neutrino sector
\begin{equation}\label{ratio Z and nu of mu3e}
    \frac{\left.\mathcal{B}(\mu\rightarrow eee)\right|_{Z'}}{\left.\mathcal{B}(\mu\rightarrow eee)\right|_{\nu}} \approx (10^{-1}\div10) \times \left(\dfrac{1}{y_\nu}\right)^4\left(\dfrac{0.5}{\bar \Delta}\right)^6\left(\dfrac{\varepsilon_\ell}{0.05}\right)^4 \left(\dfrac{\alpha_{\sscript{NP}}}{\alpha}\right)^2\left(\dfrac{\Lambda}{M_{Z'}}\right)^4\,.
\end{equation}
Here  $\alpha_\sscript{NP} = g'_3g'_{12}/4\pi$, where $g'_3,g'_{12}$ denotes the effective couplings of the $Z'$ in a 
heavy-light flavor-conserving current, while 
$\varepsilon_\ell$ is the suppression factor associated to the 
LFV couplings of the $Z'$, which is expected to be $\varepsilon_{\ell}=\mathcal{O}(0.05)$ 
(see App.~\ref{sec:Gauge and yukawa}).
Instead, for $\mu-e$ conversion in nuclei, we get
\begin{equation}\label{ratio Z and nu mu e nuclei}
    \frac{\left.\mathrm{Cr}(\mu-e,\rm{Al})\right|_{Z'}}{\left.\mathrm{Cr}(\mu-e,\rm{Al})\right|_{\nu}} \approx (10^{-1}\div10) \times \left(\dfrac{1}{y_\nu}\right)^4\left(\dfrac{0.5}{\bar \Delta}\right)^6\left(\dfrac{\varepsilon_\ell}{0.05}\right)^4 \left(\dfrac{\alpha_{\sscript{NP}}}{\alpha}\right)^2\left(\dfrac{\Lambda}{M_{Z'}}\right)^4\,.
\end{equation}
Note that the contribution coming from the neutrino sector, although loop-induced, can be comparable or even dominant with respect to the tree-level contribution from the gauge sector. This is because the loop suppression is    more than compensated by the smallness of the $\varepsilon_{\ell}$ mixing terms. 
However, the neutrino contribution becomes subleading  if $y_\nu \lesssim 0.5$.

\paragraph*{LFV $\bm{\tau}$ decays.}
The radiative decays  $\tau\rightarrow \ell\gamma$ ($\ell=e,\mu$)
are slightly enhanced with respect to $\mathcal{B}(\mu\rightarrow e\gamma)$ by a factor of $\Delta_1^{-2}$ (for $\Delta_i< 1$). However, their weaker experimental sensitivities (see Tab.~\ref{tab:main LFV bounds}) prevent them from competing with muon LFV processes. The same conclusion applies to three-body LFV tau decays. In addition, the contribution to such processes coming from the gauge sector is expected to dominate due to the unsuppressed coupling of the $Z^\prime$ to the $\tau$ lepton.
For the $\tau\rightarrow \ell'\ell\,\overline{\ell}$ decay, with $\ell,\ell'=\mu,e$, we find (see App.~\ref{sec:Gauge and yukawa})
\begin{equation}\label{tau to 3 ell decay}
    \left.\mathcal{B}(\tau\rightarrow \ell'\ell\,\overline{\ell})\right|_{Z'} \sim  10^{-7} \times \left(\dfrac{\varepsilon_\ell}{0.05}\right)^2 \left(\dfrac{\alpha_{\sscript{NP}}}{\alpha}\right)^2\left(\dfrac{\mathrm{TeV}}{M_{Z'}}\right)^4\,,
\end{equation}
which is  below current (future)  bounds 
for $M_{Z'}\gtrsim 2$ TeV (6~TeV).

\paragraph*{LFV $\bm{Z}$ decays.} The contribution coming from the neutrino sector to the decays $Z\rightarrow \ell\ell'$, with $\ell,\ell'=e,\mu,\tau$, is expected to be~\cite{Ilakovac:1994kj}
\begin{equation}
   \mathcal{B}(Z\rightarrow \ell\ell')\sim 10^{-10}\left(\dfrac{3\,\mathrm{TeV}}{\Lambda}\right)^4\,.
\end{equation}
This makes it not competitive with the other LFV observables, even assuming their future sensitivities, see Tab.~\ref{tab:main LFV bounds}. The contribution of the heavy gauge bosons is also beyond current and realistic future reach.

\paragraph*{Neutrino-less double-$\bm{\beta}$ ($\bm{0\nu2\beta}$) decays.}
In principle, a potential constraint on this framework is set by the limit on  $0\nu2\beta$ nuclear decays. Within an ISS scenario, the HNLs have a small Majorana-mass component which provides the following non-standard  contribution to the effective mass relevant for
$0\nu2\beta$ decays~\cite{Blennow:2010th}\footnote{This expression holds under the assumption that the Majorana mass matrix $\mu$ is proportional to the identity matrix. However, this approximation does not change the conclusions.}
\begin{equation}
    m_{ee}^{\rm eff}=\left|\sum_{i=1}^3U_{ei}^2m_\nu^i+p^2\sum_{k=1}^3 V_{ek}^2\dfrac{\mu}{M_k^2} \right|\equiv\left|m_{ee}+\Delta m_{ee}\right|\,,
\end{equation}
where $p\lesssim 1$ GeV is the typical momentum exchanged in the nucleus and  $m_{ee}$ is the result obtained in the minimal 
SM extensions where only Majorana masses for the left-handed neutrinos are introduced (\textit{e.g.}, in the Type-I seesaw model where the right-handed neutrinos are very heavy and decoupled).
Assuming the  conservative lower bound $M_k\gtrsim \mathcal{O}(10)$ GeV, we get
\begin{equation}
    \Delta m_{ee}\sim \mu\dfrac{p^2}{M_k^2}V_{ek}^2 \sim \mu \left(\dfrac{v}{\Lambda}\right)^2\dfrac{p^2}{M_k^2}\sim m_\nu \dfrac{p^2}{M_k^2}\ll m_{ee}\,.
\end{equation}
We thus conclude that the effect is not distinguishable from that of a minimal SM extension with massive neutrinos.

\paragraph*{PMNS non-unitarity.}
Having more than three neutrinos, the PMNS matrix in Eq.~\eqref{mixing matrices parametrization} is non-unitary.
Deviations from unitarity in the PMNS matrix are strongly constrained and are usually reported as 
bounds on $\eta\equiv|\mathbb{1}-UU^\dag|$~\cite{Antusch:2014woa}.
Within our framework, we can write 
\begin{equation}
    \eta\equiv|\mathbb{1}-UU^\dag| \sim \left(\dfrac{y_\nu v}{\Lambda}\right)^2\begin{pmatrix}
         \Delta_1^2\Delta_2^2 & \Delta_1^2 \Delta_2 & \Delta_1\Delta_2 \\ \Delta_1^2 \Delta_2 & \Delta_1^2 & \Delta_1 \\ \Delta_1 \Delta_2 & \Delta_1 & 1 
    \end{pmatrix}\,.
\end{equation}
If $\Delta_i<1$, the most stringent limit comes from $\eta_{33}$. Using the results in~\cite{Antusch:2014woa}, the bound is satisfied for $\Lambda\gtrsim2.5\,y_\nu$ TeV. Instead, if $\Delta_i>1$, then the strongest constraint follows from $\eta_{11}$ and implies  $\Lambda\gtrsim 4\,y_\nu\Delta_1\Delta_2$ TeV. In both cases, the bounds are less stringent compared to those from direct HNL searches discussed in Sect.~\ref{subsec: Direct searches}. Finally, the bounds on $\eta_{12}$ and $\eta_{22}$ stemming from PMNS non-unitarity can be safely neglected once the current bound on $\mu\rightarrow e\gamma$ is imposed.

\subsection{Implications for model building}
\label{eq:UV}

So far, we have analyzed the constraints on the parameters appearing in the neutrino Lagrangian~\eqref{mass lagrangian neutrinos} in a model-independent way, first requiring a viable light neutrino mass matrix (Sect.~\ref{subsec:range_of_param}), and then analyzing bounds from HNL searches (Sect.~\ref{subsec: Direct searches}) and low-energy processes (Sect.~\ref{subsec: LFV pheno}).
We now briefly discuss what are the implications of these constraints for the three general classes of FD models  presented in~\cite{Greljo:2024ovt}. In particular, we aim to understand which models allow the scale $\Lambda$ to be in the few TeV range, that we consider the most appealing option from a model-building point of view.

\noindent The different classes of models are identified by the choice of the subgroups of 
\begin{equation}
\mathrm{SU}(3)_\mathrm{C}\times \mathrm{SU}(2)_\mathrm{L}\times \mathrm{U}(1)_\mathrm{R}\times \mathrm{U}(1)_\mathrm{B-L}\,,
\end{equation}
that are deconstructed in flavor space. According to the analysis of 
 Ref.~\cite{Greljo:2024ovt}, there are three  options that can lead to an anarchic neutrino spectrum.  We discuss these options in detail below. In all cases, the resulting $M_R$ has the parametric form in Eq.~\eqref{M_R parametric}, while differences arise in the parametric form of $Y_\nu$.
\paragraph{Deconstruction of $\bm{\mathrm{U}(1)_\mathrm{R}}$ only.} Denoting $\phi^R_{ij}$ the link fields
   responsible for the breaking chain 
   $\mathrm{U}(1)_\mathrm{R}^3\times \mathrm{U}(1)_\mathrm{B-L} \to \mathrm{U}(1)_Y$,
the parametric expression for $Y_\nu$ reads
\begin{equation}    
Y_\nu\sim \begin{pmatrix}
        \varepsilon_1 \varepsilon_2\,  & \varepsilon_1 & 1 \\ 
        \varepsilon_1 \varepsilon_2\,  & \varepsilon_1 & 1 \\ 
        \varepsilon_1 \varepsilon_2\,  & \varepsilon_1 & 1 
    \end{pmatrix}\,,
\end{equation}
where
\begin{equation}
    \varepsilon_1\sim\eta_1\sim \dfrac{\langle\phi^R_{32}\rangle}{\Lambda_{32}}\,,\qquad  \varepsilon_2\sim\eta_2\sim \dfrac{\langle\phi^R_{21}\rangle}{\Lambda_{21}}\,.
\end{equation}
This option (corresponding to model A in~\cite{Greljo:2024ovt}) is disfavored by the LFV bounds for two main reasons.
Firstly, $Y_\nu$ is not diagonalized to the left by a unitary matrix. Barring fine-tuned cancellations, this implies the angles $\alpha_i$ are not close to the PMNS ones, effectively removing the suppression of LFV rates provided by possibly small values of $\Delta_i$. Secondly, given $\epsilon_i$ and $\eta_i$ depend on the same VEVs, it is also not easy to achieve small $\Delta_i$ in this setup.  As a result, in this case enhanced tuning is necessary to evade the strong bounds from LFV processes for 
$\Lambda\lesssim 10$~TeV
and $y_\nu =\mathcal{O}(1)$. 

\paragraph{Deconstruction of 
$\bm{\mathrm{SU}(2)_\mathrm{L} \times \mathrm{U}(1)_\mathrm{B-L}}$.} Here the parametric expression for $Y_\nu$ reads
\begin{equation}
    Y_\nu \sim \begin{pmatrix}
        \varepsilon_1 \varepsilon_2\,\, & \varepsilon_1 \varepsilon_2\eta_2\,\, & \varepsilon_1 \varepsilon_2\eta_1\eta_2 \\ \varepsilon_1\eta_2 & \varepsilon_1 & \varepsilon_1 \eta_1 \\ \eta_1\eta_2 & \eta_1 & 1
    \end{pmatrix}\,,
    \label{eq:Yv_B}
\end{equation}
where, using a self-explanatory notation for the link fields,
the parametric form of the small entries is 
\begin{equation}
    \varepsilon_1\sim \dfrac{\langle\phi_{32}^L\rangle}{\Lambda_{32}}\,,\quad \varepsilon_2\sim \dfrac{\langle\phi_{21}^L\rangle}{\Lambda_{21}}\,,\quad \eta_1\sim \dfrac{\langle\phi_{32}^{(B-L)}\rangle}{\Lambda_{32}}\,,\quad \eta_2\sim \dfrac{\langle\phi_{21}^{(B-L)}\rangle}{\Lambda_{21}}\,.
\end{equation}
In this case, one has the freedom to choose small $\Delta_i\lesssim 1$ quite naturally, as $\varepsilon_i$ and $\eta_i$ depend on VEVs of different scalars. Moreover, $Y_\nu$ is close to diagonal, hence the  
$\alpha_i$ are close to the PMNS angles. 
Setting for instance $\Delta_1=0.45$ and $\Delta_2=0.3$, which are the parameters used to make the plot in Fig.~\ref{fig:Br-mutoegamma}, 
the bounds from LFV and direct searches are satisfied  for $\Lambda\gtrsim 5$ TeV, even if $y_\nu =\mathcal{O}(1)$.

\paragraph{Deconstruction of $\bm{\mathrm{SU}(2)_\mathrm{L} \times \mathrm{U}(1)_\mathrm{R} \times \mathrm{U}(1)_\mathrm{B-L}}$.} Here the parametric expression for $Y_\nu$ reads
\begin{equation}
    Y_\nu \sim \begin{pmatrix}
        \varepsilon_1 \varepsilon_2\,\, & \varepsilon_1 \varepsilon_2\xi_R \,\, & \varepsilon_1 \varepsilon_2\xi_R^2 \\ 
        \varepsilon_1 \varepsilon_2\xi_L & \varepsilon_1    & \varepsilon_1 \xi_R \\  
        \varepsilon_1 \varepsilon_2 \xi_L^2 & \varepsilon_1 \xi_L & 1
    \end{pmatrix}\,,
\end{equation}
where 
\begin{equation}
    \varepsilon_1\sim \dfrac{\langle\phi_{32}^L\rangle\langle\phi_{32}^R\rangle}{\Lambda^2_{32}}\,,\quad \varepsilon_2\sim \dfrac{\langle\phi_{21}^L\rangle\langle\phi_{21}^R\rangle}{\Lambda^2_{21}}\,,\quad \eta_1\sim  \dfrac{\langle\phi_{32}^{(B-L)} \rangle \langle\phi_{32}^R\rangle}{\Lambda^2_{32}}\,,\quad \eta_2\sim  \dfrac{\langle\phi_{21}^{(B-L)} \rangle \langle\phi_{21}^R\rangle}{\Lambda^2_{21}}\,,
    \label{eq:Yv_C}
\end{equation}
and
\begin{equation}
   \xi_R \sim \dfrac{\langle\phi_{32}^{(B-L)}\rangle}{\langle\phi_{32}^R\rangle}
   \sim \dfrac{\langle\phi_{21}^{(B-L)}\rangle}{\langle\phi_{21}^R\rangle}\,,\qquad 
   \xi_L \sim \dfrac{\langle\phi_{32}^{(B-L)}\rangle}{\langle\phi_{32}^L\rangle}
   \sim \dfrac{\langle\phi_{21}^{(B-L)}\rangle}{\langle\phi_{21}^L\rangle}\,.
\end{equation} 
Besides a different dependence of $\varepsilon_i$ and $\eta_i$ from the model parameters, the parametric structures  in Eq.~\eqref{eq:Yv_C} and Eq.~\eqref{eq:Yv_B} share the  same virtues, hence the same considerations hold.

In summary, while minimal models based on the deconstruction of $U(1)_R$ only pose tuning problems at low $\Lambda$, the other two classes of models are phenomenologically viable. Moreover, it is worth noting that models with deconstructed $\mathrm{SU}(2)_\mathrm{L}$ are currently subject to additional constraints excluding masses of the lowest-lying $W^\prime$ states below approximately 9~TeV~\cite{Davighi:2023xqn}. Therefore, in certain regions of parameter space, the limits arising from the sector responsible for neutrino mass generation can become the most stringent bounds on the scale of the lowest layer in flavor deconstruction.

\section{Conclusion} 
\label{sec:conclusion}

We have analyzed the neutrino sector in the framework of flavor deconstruction, focusing on the inverse-seesaw realization.
This framework naturally accommodates an anarchic light-neutrino mass matrix while preserving the hierarchical flavor structure of charged fermions, and predicts heavy neutral leptons below the TeV scale.
Starting from the general parametrization of the neutrino mass matrices implied by the FD hypothesis, we identified in detail the parameter ranges consistent with neutrino oscillation data and an anarchic sterile-sector mass matrix.  

The phenomenological analysis shows that current limits from direct HNL searches and from  charged lepton flavor violation processes already set stringent bounds on the overall mass scale $\Lambda$ of the right-handed neutrinos. For $\mathcal{O}(1)$ Yukawa couplings, collider searches typically require $\Lambda \gtrsim 20~\text{TeV}$ if at least one HNL lies below the electroweak scale. However, this bound can be relaxed to the few-TeV range for heavier HNL spectra, which requires moderately suppressed Yukawa couplings, $y_\nu \lesssim 0.3$. In this regime, LFV constraints---dominated by $\mu \to e\gamma$ and $\mu$--$e$ conversion---become the leading probes.
Right now they are consistent with $\Lambda\sim$~few TeV; however, next-generation sensitivities will explore most of the parameter space with $\Lambda \lesssim 10~\text{TeV}$.  

The  phenomenological constraints from HNL searches and LFV processes have also allowed us to draw some general model-building considerations. Models based on the deconstruction of $\mathrm{SU}(2)_\mathrm{L} \times \mathrm{U}(1)_\mathrm{B-L}$ or $\mathrm{SU}(2)_\mathrm{L} \times \mathrm{U}(1)_\mathrm{R} \times \mathrm{U}(1)_\mathrm{B-L}$ gauge groups are consistent with all present bounds for $\Lambda\sim$~few TeV. On the other hand, minimal models where only the right-handed group is deconstructed require additional tuning (or heavier values of $\Lambda$). 

More generally, our analysis shows that the inverse-seesaw FD setup is a predictive  scenario linking neutrino anarchy to TeV-scale flavor dynamics, with clear signatures at colliders and LFV experiments that can be tested in the near future. 

\section*{Acknowledgment} 
\label{sec:acknowledgment}

This project has received funding by the Swiss National Science Foundation~(SNF) under contract~2000-1-240011.
This work received funding by the INFN Iniziativa Specifica APINE and by the European Union’s Horizon 2020 research and innovation programme under the Marie Sklodowska-Curie grant agreements n. 860881 – HIDDeN, n. 101086085 – ASYMMETRY. This work was also partially supported by the Italian MUR Departments of Excellence grant 2023-2027 “Quantum Frontiers” and by the European Union – Next Generation EU and by the Italian Ministry of University and Research (MUR) via the PRIN 2022 project n. 2022K4B58X – AxionOrigins. The work of N.S. is funded by the Italian Ministero dell'Università e della Ricerca through the FIS 2 project FIS-2023-01577 (DD n. 23314 10-12-2024, CUP C53C24001460001).

\appendix
\section*{Appendix} 
\label{sec:appendix}

\section{Diagonalization of the neutrino mass matrix}
\subsection{Rotation to the mass basis}\label{sec: appendix - mass basis diag}
Following \cite{Grimus:2000vj}, to diagonalize the full neutrino mass matrix encoded in the Lagrangian \eqref{mass lagrangian neutrinos}, we start by block-diagonalizing it to disentangle the light and heavy mass states by exploiting the following rotation
\begin{equation}\label{M diagonalization}
    W^\dag \begin{pmatrix}
         0 & vY_\nu & 0 \\ vY_\nu^T & 0 & M_R^T \\ 0 & M_R & \mu
    \end{pmatrix} W^*=\begin{pmatrix}
        m_\nu & \begin{matrix}
           \, 0\, & \, 0\,
        \end{matrix} \\ \begin{matrix}
            0 \\ 0
        \end{matrix} & M_h
    \end{pmatrix}=\begin{pmatrix}
        A\mu A^T & 0 & 0 \\ 0 & 0 & M_R^T \\ 0 & M_R & \mu
    \end{pmatrix}\,,
\end{equation}
where $m_\nu$ and $M_h$ being Eqs.~\eqref{m_nu as mu} and \eqref{M_h def} respectively, while
\begin{gather}
    W=\begin{pmatrix}
        1-\frac{1}{2}AA^\dag & 0 & A \\ 0 & 1 & 0 \\ -A^\dag & 0 & 1-\frac{1}{2}A^\dag A
    \end{pmatrix}\,,\qquad A=vY_\nu M_R^{-1}\,.
\end{gather}
This diagonalization is correct up to third-order corrections in $A$.\footnote{Despite the fact that $M_R$ is hierarchical, this approximation works as long as $A$ is "small". Remarkably, in such a scenario this matrix is anarchic and scales as $A\sim v/\Lambda\ll 1$} In addition, in $M_h$ the small Majorana contribution $\mu$ can be safely neglected for the phenomenology we are interested in this work. Finally, we have to diagonalize $m_\nu$ and $M_h$ using the following rotations
\begin{equation}\label{diagonalization of neutrino masses}
    V_L^\dag m_\nu V_L^*=\hat{m}_\nu=\begin{pmatrix}
        m_{\nu_1} & & \\ & m_{\nu_2} & \\ & & m_{\nu_3}
    \end{pmatrix}\,,\qquad  U_S^\dag M_R U_R=\hat{M}=\begin{pmatrix}
        M_1 & & \\ & M_2 & \\ & & M_3
    \end{pmatrix}\,.
\end{equation}
To summarize, we go to the mass basis by exploiting the following rotation
\begin{equation}
    \begin{pmatrix}
        \nu_L \\ \nu_R^c \\ s_L
    \end{pmatrix}\quad\rightarrow\quad W\begin{pmatrix}
        V_L & & \\ & U_R^* & \\ & & U_S
    \end{pmatrix}\begin{pmatrix}
        \nu_L \\ \nu_R^c \\ s_L
    \end{pmatrix}\,.
\end{equation}
The mass basis contains three light Majorana neutrinos $\nu_L^i$, $i=1,2,3$, with masses $m_{\nu_i}$ and three Heavy Neutral Leptons $n^i$ with (almost) Dirac-type masses $M_i$.

\subsection{Choice of the parameterization}\label{sec: appendix - choice of param}
To motivate the choice of our parameterization in Sect.~\ref{subsec:SM_int}, we have the freedom to rotate $Y_\nu$ or $M_R$ to be diagonal matrices by using a bi-unitary diagonalization as follows
\begin{equation}
    U_L^\dag Y_\nu U_R'=\hat{Y}_\nu=y_\nu\begin{pmatrix}
        \varepsilon_1\varepsilon_2 & & \\ & \varepsilon_1 & \\ & & 1
    \end{pmatrix}\,,\qquad U_S^\dag M_R U_R=\hat{M}=\Lambda\begin{pmatrix}
        \eta_1\eta_2 & & \\ & \eta_1 & \\ & & 1
    \end{pmatrix}\,.
\end{equation}
In principle, we cannot put both in diagonal form, but we can use the freedom to rotate $M_R\rightarrow \hat{M}$ and $Y_\nu\rightarrow \hat{Y}_\nu (U'_R{}^{\dag} U_R)$. If we then rotate also the electron Yukawa matrix as follows
\begin{equation}\label{ELECTRON YUKAWA ROTATION}
     L_E^\dag Y_e R_E=\hat{Y}_e=\begin{pmatrix}
         y_e & & \\ & y_\mu & \\ & & y_\tau
     \end{pmatrix}\,,
\end{equation}
we find the parametrization exploited in Eqs. \eqref{neutrino mass states diagonal}, \eqref{mixing matrices parametrization} and \eqref{W_parametrization}, where 
\begin{equation}\label{A from A hat }
    \mathcal{N}=L_E^\dag V_L\,,\qquad \mathcal{U}=U_L^\dag V_L\,,\qquad \hat{A}\rightarrow A=v\hat{Y}_\nu (U'_R{}^{\dag} U_R)\hat{M}^{-1}\,.
\end{equation}

There are some consequences that emerge from such a parameterization, and they are all implied by the fact that the Yukawa matrices are diagonalized by unitary matrices close to the identity thanks to their hierarchical structure in flavor-deconstructed models. Firstly, we can see that $\mathcal{N}=\mathcal{U}+\mathcal{O}(\varepsilon_1,\varepsilon_2)$. This condition translates in the range of the angles that parameterize $\mathcal{U}$. Secondly, the PMNS matrix is almost fully reproduced by $V_L$, which is the matrix that diagonalizes $m_\nu$ (see Eq. \eqref{diagonalization of neutrino masses}). This observation will be crucial in determining the possible range for $\Delta_i$. Lastly, $\hat{A}$ is not exactly diagonal, but it gets a small correction. We will discuss this in more detail below. 

Finally, for completeness, we write the Majorana mass matrix of the sterile fermions $\mu$ in terms of such parameterization
\begin{equation}
    \mu=W^{-1}\hat{m}_\nu (W^{-1})^T\,.
\end{equation}

\subsection{Non-diagonal corrections to matrix A}\label{sec: appendix - corr to A}
In this parameterization we have that 
\begin{equation}
    A=v\hat{Y}_\nu (U'_R{}^{\dag} U_R)\hat{M}^{-1}=v\hat{Y}_\nu\left[\mathbb{1}+\mathcal{O}(\varepsilon_i,\eta_i)\right]\hat{M}^{-1}\equiv \hat{A}+\delta A\,.
\end{equation}
To represent $3\times 3$ unitary matrices (neglecting the complex phases), we exploit the following parameterization
\begin{equation}
    U[\theta_1,\theta_2,\theta_3]=\begin{pmatrix}
            1 & 0 & 0 \\ 0 & \cos{\theta_1} & \sin{\theta_1} \\ 0 & -\sin{\theta_1} & \cos{\theta_1}
        \end{pmatrix}\begin{pmatrix}
           \cos{\theta_2} & 0 & \sin{\theta_2} \\ 0 & 1 & 0 \\ -\sin{\theta_2} & 0 & \cos{\theta_2}
        \end{pmatrix}\begin{pmatrix}
            \cos{\theta_3} & \sin{\theta_3} & 0 \\ -\sin{\theta_3} & \cos{\theta_3} & 0 \\ 0 & 0 & 1
        \end{pmatrix}\,.
\end{equation}
By knowing the hierarchical structure of $M_R$, see Eq. \eqref{M_R parametric}, it is possible to show that we expect the angles parameterizing $U_R$ to be of order $\theta_1\sim \eta_1$, $\theta_3\sim \eta_2$, $\theta_2\sim \eta_1\eta_2$. The same is expected for $U_R'$. Hence we can observe that, for $\eta_i\gtrsim \varepsilon_i$ (or equivalently $\Delta_i\lesssim 1$), 
\begin{equation}
    \delta A\sim \dfrac{y_\nu v}{\Lambda}\begin{pmatrix}
             0 &  0 &  0 \\  \Delta_1 &  0 &  0 \\  1 &  1 &  0
    \end{pmatrix}\,,
\end{equation}
while for $\eta_i\lesssim \varepsilon_i$ (or equivalently $\Delta_i\gtrsim 1$), 
\begin{equation}
    \delta A\sim \dfrac{y_\nu v}{\Lambda}\begin{pmatrix}
             0 &  0 &  0 \\  \Delta_1\Delta_2 &  0 &  0 \\  \Delta_1\Delta_2 &  \Delta_1 &  0
    \end{pmatrix}\,.
\end{equation}

Nevertheless, the phenomenology depends on $W$ in Eq. \eqref{W_parametrization}. Since $U_S=\mathcal{O}(1)$, we notice that, for $\Delta_i\lesssim 1$,
\begin{equation}
    AU_S^\dag \sim \frac{y_\nu v}{\Lambda}\begin{pmatrix}
        \Delta_1\Delta_2 & \Delta_1\Delta_2 & \Delta_1\Delta_2 \\ \Delta_1 & \Delta_1 & \Delta_1 \\ 1 & 1 & 1
    \end{pmatrix}\sim \hat{A}U_S^\dag\,.
\end{equation}
Therefore, even if one adds three more parameters to parameterize the non-diagonal entries of $A$, it does not effectively change the phenomenology. Remarkably, we have checked that this is indeed the case from numeric simulations. On the other hand, for $\Delta_i\gtrsim 1$,
\begin{equation}
    AU_S^\dag \sim \frac{y_\nu v}{\Lambda}\begin{pmatrix}
        \Delta_1\Delta_2 & \Delta_1\Delta_2 & \Delta_1\Delta_2 \\  \Delta_1\Delta_2 & \Delta_1\Delta_2 & \Delta_1\Delta_2 \\  \Delta_1\Delta_2 & \Delta_1\Delta_2 & \Delta_1\Delta_2
    \end{pmatrix}\,.
\end{equation}
In this latter case possible non-diagonal entries in $A$ could lead to an enhancement in the coupling of the HNLs with $\mu$ and $\tau$ by a factor of order $\Delta_2$ and $\Delta_1\Delta_2$, respectively. In any case, since the case $\Delta_i\gtrsim 1$ points towards NP scales $\Lambda\gg 10$ TeV (see Sect. \ref{subsec: Direct searches}), this scenario is already not interesting in the context of FD. 

\section{Contribution from gauge and Yukawa sectors to LFV processes}
\label{sec:Gauge and yukawa}

The gauge and Yukawa sectors of flavor-deconstructed models already generate NP contributions to LFV processes even without including neutrinos. Hereafter, we want to quantify the predictions from such sectors and compare them with the ones coming from the neutrino sector.

\subsection{Generic features of LFV in flavor deconstruction}

Although model dependent, we can still infer some general features of LFV effects as arising from the gauge and Yukawa sectors of flavor-deconstructed models. In particular, the gauge sector contains at least one massive neutral gauge boson that couples non-universally with the fermions of the theory as follows:
\begin{equation}
    \mathcal{L}\supset |D_\mu H|^2+Z'_\mu J_{Z'}^\mu+\dfrac{1}{2}M_{Z'}^2Z_\mu'Z'^\mu\,,
\end{equation}
where
\begin{equation}
    J_{Z'}^\mu=g_\sscript{NP}\sum_{\psi}Q^\psi_{Z'}\overline{\psi}\gamma^\mu \psi\,,\qquad D_\mu H\supset -i g_{\sscript{NP}} Q^H_{Z'}Z'_\mu H\,,
\end{equation}
where $Q^{\psi,H}_{Z'}$ are non-universal gauge charges. In FD the first and second generations are deconstructed at much higher energies than the third one. Therefore, the leading effects to phenomenological observables stem from the massive vector boson related to the third generation with a mass or order $M_{Z'}\sim g_{\sscript{NP}} \langle\phi\rangle$, where $\phi$ is some NP heavy scalar field. Moreover, the Yukawa sector of the theory can be schematically written as
\begin{equation}
    -\mathcal{L}\supset y_{33}\overline{\ell}_3 H e_3+\sum_{j=1,2}\sum_{\alpha=\mathrm{heavy}}Y_{j\alpha}\overline{\ell}_j H E_\alpha+\left[\sum_{\beta=\mathrm{heavy}}Y'_{\beta i}\overline{E}_\beta \phi e_2+1^{\mathrm{st}}\,\mathrm{ generation}\right]\,,
\end{equation}
where, for simplicity, we just focused on the electron Yukawa couplings. In particular, only the $33$ SM Yukawa interactions are generated at the renormalizable level, along with other Yukawa-like renormalizable interactions between light and heavy vectorlike fermions $E_\alpha$. 

After the spontaneous symmetry breaking of the UV gauge group, mass mixing between heavy and light fermions are generated. Going to the mass basis, fermions are rotated by the following mixing matrix
\begin{equation}\label{mass mixing rotation}
    U_\psi=\begin{pmatrix}
        \mathbb{1}-\frac{1}{2}\mathcal{E}_\psi^\dag\mathcal{E}_\psi & \mathcal{E}^\dag_\psi \\ -\mathcal{E}_\psi & \mathbb{1}-\frac{1}{2}\mathcal{E}_\psi\mathcal{E}^\dag_\psi
    \end{pmatrix}\,,\qquad \mathcal{E}_\psi\sim \langle \phi \rangle/M_F\,,
\end{equation}
where $\psi=q,\ell,u,d,e$, while $M_F$ is the mass of the heavy fermions. In such a way, one can generate the Yukawa couplings for the light generations that are schematically given by
\begin{equation}
    (Y_{f})_{ji}\sim (Y\cdot \mathcal{E}_\psi)_{ji}\,,
\end{equation}
where $f=u,d,e$ and $(j,i)\neq (3,3)$, while $\mathcal{E}_\psi$ has to account for the hierarchical structure of the SM Yukawa matrices, which happens when the mass-scale of the vectorlike fermions is larger than the vev of the link fields $M_F\gg \langle \phi\rangle$. After the electroweak symmetry breaking, we obtain the SM fermion mass matrices which are diagonalized by two unitary matrices as follows
\begin{equation}
    U_L^\dag Y_f U_R=\hat{Y}_f=\mathrm{diag}(y_1,y_2,y_3)\,,\qquad U_{L/R}=\mathbb{1}+\Delta_\psi\,,\qquad \Delta_\psi\sim \langle \phi \rangle/M_F\,.
\end{equation}
In particular, the unitary rotation matrices are close to the identity as $Y_f$ is hierarchical and almost diagonal, with off-diagonal entries being of the same order of $\mathcal{E}_\psi$ in Eq. \eqref{mass mixing rotation}. Such rotations affect the fermionic currents which develop  flavor-violating entries that are suppressed by powers of $\mathcal{E}_\psi$ and $\Delta_\psi$. In particular, we obtain 
\begin{equation}
    J_{Z'}^\mu\supset g_{\sscript{NP}}\sum_{\psi=q,\ell,u,d,e}\,\sum_{p,r=1}^3\overline{\psi}_p\gamma^\mu\psi_r P_{pr}(\psi)\,,
\end{equation}
where
\begin{equation}
\begin{split}
    P_{pr}(\psi)=&Q_{Z'}(\psi_p)\delta_{pr}+\sum_{\alpha= \mathrm{heavy}}Q_{Z'}(\psi_\alpha)(\mathcal{E}_\psi)^*_{\alpha p}(\mathcal{E}_\psi)_{\alpha r}
    -\dfrac{1}{2}\Big[Q_{Z'}(\psi_p)+Q_{Z'}(\psi_r)\Big](\mathcal{E}_\psi^\dag\mathcal{E}_\psi)_{pr}\\
    &+Q_{Z'}(\psi_p)(\Delta_\psi)_{pr}+Q_{Z'}(\psi_r)(\Delta_\psi)^*_{rp}+\sum_{k=\mathrm{light}}Q_{Z'}(\psi_k)(\Delta_\psi)_{kp}^*(\Delta_\psi)_{kr}\,.
\end{split}
\end{equation}
Hence, flavor-violating entries involving light-generations and the third one (i.e. $p=3$, $r=1,2$) are linearly suppressed by $\Delta_\psi$, while the strength of flavor violation among light generations (i.e. $p,r=1,2$) are quadratically suppressed by either $\mathcal{E}_\psi^2$ or $\Delta_\psi^2$.\footnote{The quadratic dependence on $\Delta_\psi^2$ results from the equality $Q_{Z'}(\psi_1)=Q_{Z'}(\psi_2)$ and by the unitarity condition $\Delta_\psi+\Delta_\psi^\dag=-\Delta_\psi^\dag\Delta_\psi$.} Hereafter, for simplicity, we adopt the notation (here $i=1,2$)
\begin{equation}\label{varepsilon mixing angles behaviour}
    g_{\sscript{NP}}P_{3i}(\ell_{L/R})\sim g'_3\varepsilon_{\ell_i}\,,\qquad g_\sscript{NP} P_{12}(\ell_{L/R})\sim g'_{3}\varepsilon_e\varepsilon_\mu\,,\qquad g_\sscript{NP}P_{ii}(\ell_{L/R})\sim g'_{12}\,,
\end{equation}
where $\varepsilon_\ell \sim \mathcal{E}_\psi,\,\Delta_\psi$, while $g'_3,g'_{12}$ account for the different gauge charges $Q_{Z'}$ of the $Z'$ boson to different fermion generations, as in FD usually the coupling strength with the third generation is enhanced with respect to the light ones. As a benchmark, we can further assume that  $\varepsilon_\ell\sim m_e/m_\mu \sim 0.05$. 

After integrating out the heavy DOFs, the relevant SMEFT operators for LFV processes are given by~\footnote{Dipole operators are not discussed in our analysis as they are generated at loop-level and therefore subleading with respect to the tree-level effects, as we explicitly checked.}
\begin{subequations}
    \begin{gather}
        (Q_{H\psi}^{(1)})_{pr}=H^\dag i \overleftrightarrow{D}_\mu H (\overline{\psi}_p \gamma^\mu \psi_r)\,,\\
        (Q_{\psi\psi'})_{pqrs}=(\overline{\psi}_p\gamma^\mu\psi_q)(\overline{\psi}'_r\gamma_\mu \psi'_s)\,,
    \end{gather}
\end{subequations}
where $\psi,\psi'=\ell,q,u,d,e$. The corresponding Wilson coefficients read~\footnote{Here $\Lambda$ is a generic NP scale not to be confused with the $\Lambda$ used in the main text.}
\begin{subequations}\label{SMEFT tree-level}
    \begin{gather}
        (C^{(1)}_{H\psi})_{pr}/\Lambda^2=-\frac{1}{M_{Z'}^2}(g_{\sscript{NP}})^2Q_{Z'}(H)P_{pr}(\psi)\,,\\
        (C_{\psi\psi'})_{pqrs}/\Lambda^2=-\dfrac{1}{\mathcal{S}_{\psi,\psi'}M_{Z'}^2}(g_{\sscript{NP}})^2 P_{pq}(\psi) P_{rs}(\psi')\,,
    \end{gather}
\end{subequations}
where $\mathcal{S}_{\psi,\psi'}$ is a symmetry factor equal to $1$ if $\psi\neq \psi'$, or $2$ if $\psi=\psi'$. 
\begin{figure}
    \centering
    \includegraphics{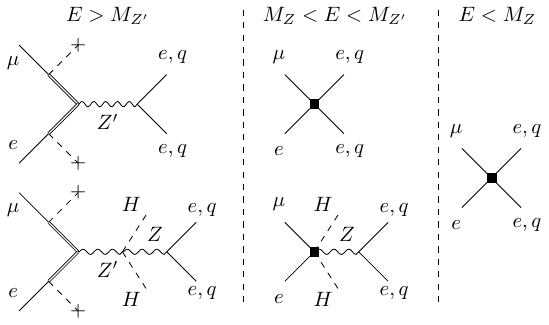}
    \caption{Tree-level contributions to Wilson coefficients of 
    the operators mediating $\mu eee$ and $\mu eqq$ processes at different energy scales.}
    \label{fig:mu3e_scales}
\end{figure}
To account for LFV muon and tau decays, the next step is to match the SMEFT into the LEFT which takes place below the electroweak scale. LEFT operators can arise directly from the $\psi^4$ operators of the SMEFT, or indirectly through the $\psi^2 H^2 D$ operators. In the latter case, an effective mixing between the $Z_\mu$ and $Z'_\mu$ heavy boson induces modifications to the $Z_\mu$ couplings with light fermions, potentially introducing flavor-violating interactions that are not present in the Fermi theory. Schematically, this is represented in Fig.~\ref{fig:mu3e_scales}. 

\subsection{LFV processes}
Now we can provide some estimates for the LFV processes considered in Sect.~\ref{subsec: LFV pheno}.

\paragraph*{LFV $\bm{\mu}$ decays.} The prediction to $\mu\rightarrow eee$ reads \cite{Kuno:1999jp,Crivellin:2013hpa}
\begin{equation}
    \mathcal{B}(\mu\rightarrow eee)=\dfrac{1}{4G_F^2 M_{Z'}^4}\left(|C_{VLL}|^2+|C_{VRR}|^2+\frac{1}{2}|C_{VRL}|^2+\frac{1}{2}|C_{VLR}|^2\right)\,,
\end{equation}
where
\begin{subequations}
    \begin{align}
    C_{VLL}&=(2s_W^2-1)(C_{H\ell}^{(1)})_{e\mu}+(C_{\ell\ell})_{e\mu ee}\,,\\
    C_{VRR}&=2s_W^2(C_{He})_{e\mu}+(C_{ee})_{e\mu ee}\,,\\
    C_{VLR}&=2s_W^2(C_{H\ell}^{(1)})_{e\mu}+(C_{\ell e})_{e\mu ee}\,,\\
    C_{VRL}&=(2s_W^2-1)(C_{He})_{e\mu}+(C_{\ell\ell})_{eee\mu}\,.
\end{align}
\end{subequations}
Combining the Wilson coefficients computed in Eq.~\eqref{SMEFT tree-level} together with Eq.~\eqref{varepsilon mixing angles behaviour}, we find Eq.~\eqref{ratio Z and nu of mu3e}. The prediction to $\mu-e$ conversion in nuclei reads~\cite{Kuno:1999jp,Crivellin:2013hpa}
\begin{equation}
    {\mathrm{Cr}}(\mu-e,{\rm{N}})=\dfrac{m_\mu^5}{\Gamma_{\mathrm{capt}}M_{Z'}^4}\left(\left|g_{LV}^p V^p+g_{LV}^n V^n\right|^2+\left|g_{RV}^p V^p+g_{RV}^n V^n\right|^2\right)\,,
\end{equation}
where
\begin{subequations}
    \begin{align}
    g_{LV}^p&=(4s_W^2-1)(C_{H\ell}^{(1)})_{e\mu}-2(C_{\ell q}+C_{\ell u})_{e\mu uu}-(C_{\ell q}+C_{\ell d})_{e\mu dd}\,,\\
    g_{LV}^n&=(C_{H\ell}^{(1)})_{e\mu}-(C_{\ell q}+C_{\ell u})_{e\mu uu}-2(C_{\ell q}+C_{\ell d})_{e\mu dd}\,,\\
    g_{RV}^p&=(4s_W^2-1)(C_{H e})_{e\mu}-2(C_{e q}+C_{eu})_{e\mu uu}-(C_{eq}+C_{ed})_{e\mu dd}\,,\\
    g_{RV}^n&=(C_{H e})_{e\mu}-(C_{e q}+C_{eu})_{e\mu uu}-2(C_{eq}+C_{ed})_{e\mu dd}\,.
\end{align}
\end{subequations}
In a completely analogous way, by using the Al-nuclear factors $V^p\approx V^n\approx 0.02$ and $\Gamma_{\mathrm{capt}}\approx 5\cdot 10^{-22}$ TeV \cite{Heinz:2024cwg}, we find Eq.~\eqref{ratio Z and nu mu e nuclei}.

\paragraph*{LFV $\bm{\tau}$ decays.} The prediction to $\tau\rightarrow \ell'\ell\overline{\ell}$, with $\ell,\ell'=\mu,e$, reads \cite{Crivellin:2013hpa}
\begin{equation}
    \mathcal{B}(\tau\rightarrow \ell'\ell\,\overline{\ell})=\dfrac{1}{4G_F^2 M_{Z'}^4}\left(|C_{VLL}|^2+|C_{VRR}|^2+\frac{1}{2}|C_{VRL}|^2+\frac{1}{2}|C_{VLR}|^2\right)\,,
\end{equation}
where
\begin{subequations}
    \begin{align}
    C_{VLL}&=(2s_W^2-1)(C_{H\ell}^{(1)})_{\ell\tau}+(C_{\ell\ell})_{\ell\tau\ell'\ell'}\,,\\
    C_{VRR}&=2s_W^2(C_{He})_{\ell\tau}+(C_{ee})_{\ell\tau\ell'\ell'}\,,\\
    C_{VLR}&=2s_W^2(C_{H\ell}^{(1)})_{\ell\tau}+(C_{\ell e})_{\ell\tau\ell'\ell'}\,,\\
    C_{VRL}&=(2s_W^2-1)(C_{He})_{\ell\tau}+(C_{\ell\ell})_{\ell\ell'\ell'\tau}\,.
\end{align} 
\end{subequations}
Combining the Wilson coefficients computed in Eq.~\eqref{SMEFT tree-level} together with Eq.~\eqref{varepsilon mixing angles behaviour}, we find Eq.~\eqref{tau to 3 ell decay}.

\paragraph*{LFV $\bm{Z}$ decays.} The prediction to $Z\rightarrow \tau\ell$, with $\ell=e,\mu$, reads~\cite{Crivellin:2013hpa}
\begin{equation}
    \mathcal{B}(Z\rightarrow\tau\ell)=\dfrac{1}{96G_F^2 M_{Z'}^4}\dfrac{\alpha M_Z}{\Gamma_Z s_W^2 c_W^2}\left(\left|(C_{H\ell}^{(1)})_{\tau\ell}+(C_{H\ell}^{(3)})_{\tau\ell}\right|^2+\Big|(C_{He})_{\tau\ell}\Big|^2\right)\,.
\end{equation}

\bibliographystyle{JHEP}
\bibliography{Reference.bib}{}

\end{document}